\pdfoutput=1 
\documentclass[useAMS,usenatbib,usegraphicx]{mn2e}
\usepackage{aas_macros}
\usepackage{amsmath}
\usepackage{amssymb}
\usepackage{mathbbol}
\usepackage{color}

\newcommand{\kms}{\ensuremath{\mathrm{km}\,\mathrm{s}^{-1}}}

\title[The new satellites: VPOS alignment \& velocity predictions]{The new Milky Way satellites: alignment with the VPOS and predictions for proper motions and velocity dispersions}
\author[Pawlowski, McGaugh \& Jerjen]{Marcel S. Pawlowski$^{1}$\thanks{E-mail:
marcel.pawlowski@case.edu}, Stacy S. McGaugh$^{1}$, Helmut Jerjen$^{2}$ \\
$^{1}$Department of Astronomy, Case Western Reserve University, 10900 Euclid Avenue, Cleveland, OH 44106, USA\\
$^{2}$Research School of Astronomy and Astrophysics, Australian National University, Mt Stromlo Observatory,\\ Cotter Rd., Weston ACT 2611, Australia
}
\begin{document}

\date{Accepted ??? Received ???; in original form ???}

\pagerange{\pageref{firstpage}--\pageref{lastpage}} \pubyear{2015}

\maketitle

\label{firstpage}

\begin{abstract}
The evidence that stellar systems surrounding the Milky Way (MW) are distributed in a Vast Polar Structure (VPOS) may be observationally biased by satellites detected in surveys of the northern sky. The recent discoveries of more than a dozen new systems in the southern hemisphere thus constitute a critical test of the VPOS phenomenon. We report that the new objects are located close to the original VPOS, with half of the sample having offsets less than 20\,kpc. The positions of the new satellite galaxy candidates are so well aligned that the orientation of the revised best-fitting VPOS structure is preserved to within 9 degrees and the VPOS flattening is almost unchanged (31\,kpc height). Interestingly, the shortest distance of the VPOS plane from the MW center is now only 2.5 kpc, indicating that the new discoveries balance out the VPOS at the Galactic center. The vast majority of the MW satellites are thus consistent with sharing a similar orbital plane as the Magellanic Clouds, confirming a hypothesis proposed by Kunkel \& Demers and Lynden-Bell almost 40 years ago. We predict the absolute proper motions of the new objects assuming they orbit within the VPOS. Independent of the VPOS results we also predict the velocity dispersions of the new systems under three distinct assumptions: that they (i) are dark-matter-free star clusters obeying Newtonian dynamics, (ii) are dwarf satellites lying on empirical scaling relations of galaxies in dark matter halos, and (iii) obey MOND.
\end{abstract}


\section{Introduction}
\label{sect:intro}

\begin{table*}
\begin{minipage}{180mm}
 \caption{Compilation of observed properties of the newly detected MW satellite objects}
 \label{tab:obsprop}
 \begin{center}
 \begin{tabular}{@{}llcccccccccc}
 \hline 
Object & suggested names & Type & $\alpha$ & $\delta$ &  m -- M &  $r_{\sun}$ & M$_V$ & $r_{1/2}$ & $r_{1/2}$ & Ref. \\
 &   &  & [$^{\circ}$] & [$^{\circ}$] & [mag] & [kpc] & [mag] & [arcmin] & [pc]  &  \\
 \hline
Kim 1  &  Kim 1           & CC &  332.92  & 7.03 &  $16.5 \pm 0.1$ &  $19.8 \pm 0.9$  & $0.3 \pm 0.5$  & $1.2 \pm 0.1$ & $6.9 \pm 0.6$   & (1)\smallskip\\
Ret II & DES J0335.6-5403  & U  &  53.92  & -54.05 &  17.5 &   32  & $-3.6 \pm 0.1$  & $3.8^{+1.0}_{-0.6}$ & $35^{+9}_{-5}$   & (2)\\
& Reticulum 2              &    &  53.93  & -54.05 &  17.4 &   30  & $-2.7 \pm 0.1$  & $3.7 \pm 0.2$ & $32 \pm 1$   & (3)\smallskip\\
Lae II & Triangulum II      & U  &  33.32  & 36.18  &  $17.4 \pm 0.1$ &   30  & $-1.8 \pm 0.5$  & $3.9^{+1.1}_{-0.9}$ & $34^{+9}_{-8}$ &   (4)\smallskip\\
Tuc II & DES J2251.2-5836  & UD & 343.06  & -58.57 &  18.8 &   58  & $-3.9 \pm 0.2$  & $7.2 \pm 1.8$ & $120 \pm 30$    & (2)\\
& Tucana 2                 &    & 342.97  & -58.57 &  19.2 &   69  & $-4.4 \pm 0.1$  & $9.9 \pm 1.4$ & $199 \pm 28$    & (3)\smallskip\\
Hor II & Horologium II     & U  &  49.13  & -54.14 &  19.5 &   78  & $-2.6^{+0.2}_{-0.3}$  & $2.09^{+0.44}_{-0.41}$ & $47 \pm 10$   & (5)\smallskip\\
Hor I & DES J0255.4-5406   & U  &  43.87  & -54.11 &  19.7 &   87  & $-3.5 \pm 0.3$  & $2.4^{+3.0}_{-1.2}$ & $60^{+76}_{-30}$   & (2)\\
& Horologium 1             &    &  43.88  & -54.12 &  19.5 &   79  & $-3.4 \pm 0.1$  & $1.3 \pm 0.2$ & $30 \pm 3$   & (3)\smallskip\\
Phe II & DES J2339.9-5424  & U & 354.99  & -54.41 &  19.9 &   95  & $-3.7 \pm 0.4$  & $1.20 \pm 0.6$ & $33^{+20}_{-11}$    & (2)\\
& Phoenix 2                &    & 355.00  & -54.41 &  19.6 &   83  & $-2.8 \pm 0.2$  & $1.1 \pm 0.2$ & $27 \pm 5$   & (3)\smallskip\\
Eri III & DES J0222.7-5217 & U  &  35.69  & -52.28 &  19.9 &   95  & $-2.4 \pm 0.6$  & $0.4^{+0.3}_{-0.2}$ & $11^{+8}_{-5}$    & (2)\\
& Eridanus 3               &    &  35.69  & -52.28 &  19.7 &   87  & $-2.0 \pm 0.3$  & $0.7 \pm 0.3$ & $18 \pm 8$   & (3)\smallskip\\
Kim 2 &   Kim 2         & CC & 317.21  & -51.16 &  $20.1 \pm 0.1$ & $104.7 \pm 4.1$ &  $-1.5 \pm 0.5$ &$0.42 \pm 0.02 $& $12.8 \pm 0.6$   & (6)\\
       & DES J2108.8-5109  &    & 317.20  & -51.16 &  19.2 &   69  & $-2.2 \pm 0.5$  & $0.6 \pm 0.1$ & $12 \pm 2$   & (2)\\
& Indus 1                  &    & 317.20  & -51.17 &  20.0 &  100  & $-3.5 \pm 0.2$  & $1.4 \pm 0.4$ & $39 \pm 11$    & (3)\smallskip\\
Gru I & Grus 1             & U  & 344.18  & -50.16 &  20.4 &  120  & $-3.4 \pm 0.3$  & $2.0 \pm 0.7$ & $70 \pm 23$    & (3)\smallskip\\
Pic I & DES J0443.8-5017   & U  &  70.95  & -50.28 &  20.5 &  126  & $-3.7 \pm 0.4$  & $1.2^{+4.2}_{-0.6}$ & $43^{+153}_{-21}$ & (2)\\
& Pictoris 1               &    &  70.95  & -50.28 &  20.3 &  114  & $-3.1 \pm 0.3$  & $0.9 \pm 0.2$ & $31 \pm 7$     & (3)\smallskip\\
Hyd II & Hydra II          & UD & 185.44  & -31.98 &  $20.64 \pm 0.16$  & $134 \pm 10$ & $-4.8 \pm 0.3$ & $1.7^{+0.3}_{-0.2}$ & $68 \pm 11$ &   (7)\smallskip\\
Peg III & Pegasus III      & UD & 336.10  &  5.41  &  $21.56 \pm 0.20$ &  $205 \pm 20$ & $-4.1 \pm 0.5$ & $1.85 \pm 0.10$     & $110 \pm 6$ & (8)\smallskip\\
Eri II & DES J0344.3-4331  & UD &  56.09  & -43.53 &  22.6 &  330  & $-7.4 \pm 0.1$  & $1.6 \pm 0.6$ & $155 \pm 54$    & (2)\\
& Eridanus 2               &    &  56.09  & -43.53 &  22.9 &  380  & $-6.6 \pm 0.1$  & $1.6 \pm 0.1$ & $172 \pm 12$    & (3)\\
 \hline
 \end{tabular}
 \end{center}
 \small \medskip
Properties of the newly discovered MW satellite objects collected from the literature, the reference from which the data of a given row has been collected is indicated in the last column (labelled ``Ref.'').
Type can be CC for confirmed star cluster, UD for unconfirmed dwarf galaxy candidate and U for unclassified (unconfirmed star cluster or ultra-faint dwarf galaxy candidate). The other parameters are
$\alpha$\ and $\delta$\ for right ascension and declination of the object, m -- M for its distance modulus, $r_{\sun}$\ for the Heliocentric distance in kpc, M$_V$\ for the absolute v-band magnitude, and $r_{1/2}$\ for the two-dimensional half light radius in both arc minutes and parsec. 
DES15 find Ret\,II and Eri\,II to have a significant ellipticity $\epsilon$\ and report their half-light radii as measured along the semi-major axis. We report and use the circularized $r_{1/2}$, which is obtained by multiplying the semi-major axis ellipticity with $\sqrt{1 - \epsilon}$.
Uncertainties are given if they are specified in the original publication. \\
Reference: (1): \citet{KimJerjen2015}, (2): DES15, (3): \citet{Koposov2015}, (4): \citet{Laevens2015}, (5): \citet{Kim2015c}; (6): \citet{Kim2015}, (7): \citet{Martin2015}, (8): \citet{Kim2015b}.
\end{minipage}
\end{table*}

\begin{figure*}
 \centering
 \includegraphics[width=180mm]{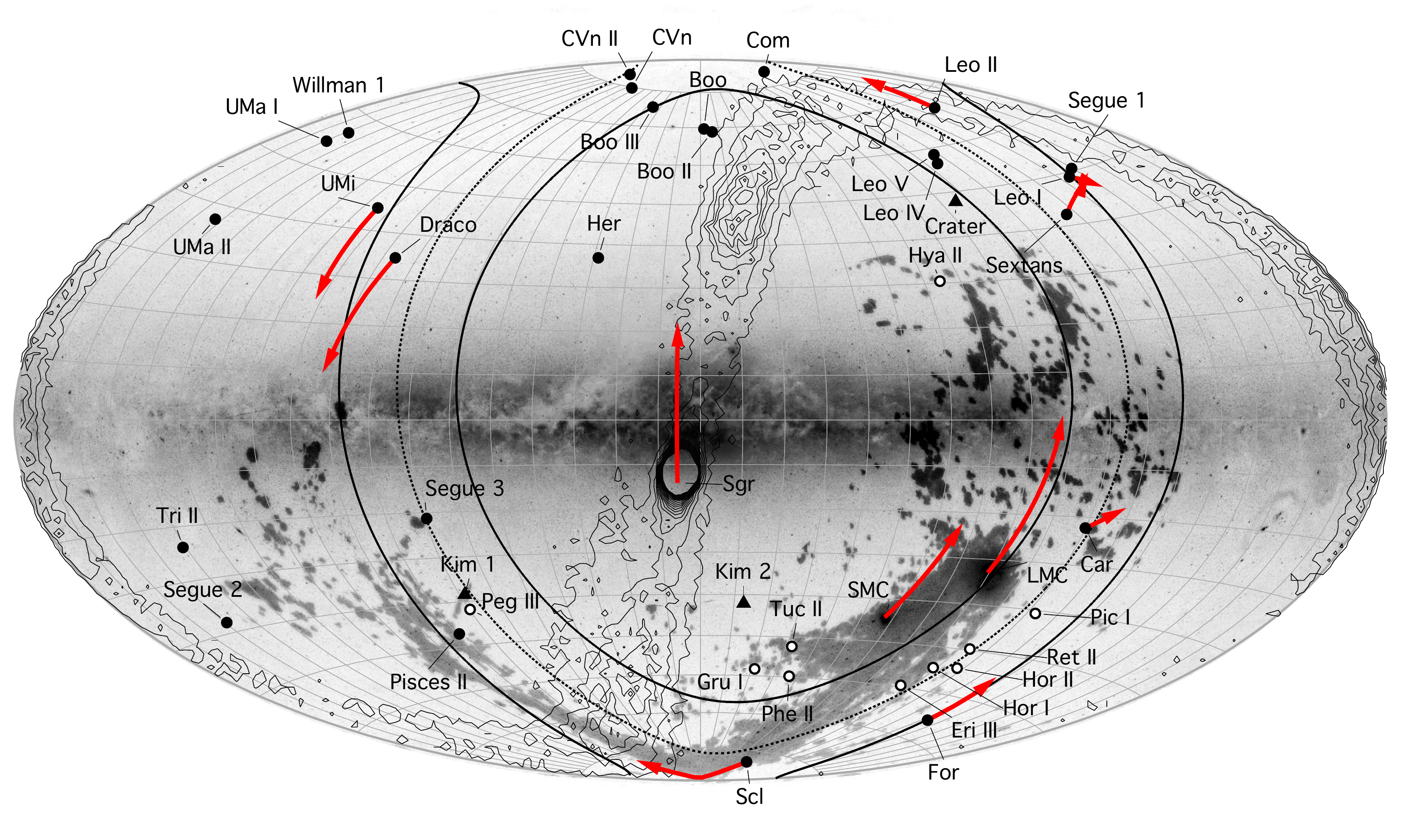}
 \caption{
 All-sky plot illustrating the positions of the MW satellites relative to the MW (inverted image in the background of the plot). Previously known satellite galaxies are plotted as filled black points, recently discovered satellite objects as white points and star clusters as filled black triangles. 
 The velocities of the 11 classical satellites are indicated by the red arrows, which are the projected 3D velocity vectors determined from the measured line-of-sight velocities and proper motions as compiled in \citet{PawlowskiKroupa2013}. 
 The dotted line indicates the orientation of the VPOS+new fit. The rms height of the VPOS is illustrated with two solid lines, which show the intersection of the VPOS+new of height 31\,kpc with a sphere of 100\,kpc radius. The majority of MW satellites fall close to the VPOS, as does the Magellanic Stream (dark patches). Note that the parallax effect due to the offset of the Sun from the Galactic center (and thus the best-fit VPOS plane) and the intrinsic thickness of the VPOS increase the scatter of the satellites around the VPOS great circle. The proper motions indicate that eight of the 11 classical satellites are consistent with co-orbiting in the VPOS, while Sculptor is counter-orbiting within the VPOS and Sagittarius is on an orbit perpendicular to both the MW and the VPOS, as also traced by the Sagittarius stream (contour lines indicate particle densities for the simulated stream from \citealt{LawMajewski2010}).
The background image is by \citet{Nidever2010}, NRAO/AUI/NSF and Meilinger, Leiden-Argentine-Bonn Survey, Parkes Observatory, Westerbork Observatory, Arecibo Observatory (see http://www.nrao.edu/pr/2010/magstream/).
 }
 \label{fig:ASP}
\end{figure*}

It has been known for almost 40 years now that the satellite galaxies of the Milky Way (MW) preferentially 
lie along a great circle in the sky which passes almost through the Galactic poles. The MW satellites are 
thus distributed in a flattened, polar plane. From the beginning it was noticed that this plane contains the 
Magellanic Clouds and follows the Magellanic stream. Hence, \citet{KunkelDemers1976} and \citet{LyndenBell1976} 
 termed this phenomenon the ``Magellanic plane''. As more MW satellites were discovered, e.g., Sextans and those dwarf galaxies 
 in the footprint of the Sloan Digital Sky Survey, they were found to lie close to this plane too \citep{Kroupa2005, Metz2009}.

With the inclusion of other types of MW halo objects this plane was subsequently called the 
Vast Polar Structure (VPOS) of the MW \citep{Pawlowski2012}. It was found that young halo globular clusters, 
which are hypothesised to have formed in external dwarf galaxies that
were accreted into the Galactic potential well, follow the same polar distribution \citep{Keller2012,Pawlowski2012}.
Even the stellar streams of some disrupting MW satellites (both galaxies and star clusters) have been found to 
align with the VPOS \citep{Pawlowski2012,PawlowskiKroupa2014}. The orbital directions of the 11 classical satellites, 
deduced from their proper motions (PMs), further indicate that most satellites co-orbit in the plane. The VPOS is  
thus not only a spatial but also a rotating structure \citep{PawlowskiKroupa2013}.

\citet{Kroupa2005} were the first to argue that the narrow, polar alignment of the 11 brightest ``classical'' 
MW satellites is in conflict with the typical distribution of dark matter sub-halos in $\Lambda$CDM. 
Whether the positional alignment of these 11 satellite galaxies is problematic for $\Lambda$CDM or not has 
been challenged since then \citep[e.g.][]{Deason2011,Libeskind2005,Libeskind2009,Wang2013,Zentner2005}. Taking into 
account the limitation of detecting satellite galaxies close to the Galactic plane, the significance of the planar arrangement 
of the 11 satellites is between 99.4 and 99.9\,per cent \citep{Pawlowski2014,PawlowskiMcGaugh2014b}. 
Due to minor anisotropies present in dark matter sub-halo systems around MW equivalents in simulations, 
it has been found that the VPOS flattening alone can be reproduced by 0.5 to 6\,per cent of $\Lambda$CDM systems 
\citep{Wang2013,Pawlowski2014,PawlowskiMcGaugh2014b}. However, once the kinematic 
correlation is being taken into account, structures like the VPOS are extraordinarily rare in 
cosmological simulations ($< 0.1$\,per cent, \citealt{Pawlowski2014}; \citealt{PawlowskiMcGaugh2014b}). 
Planar satellite arrangements sometimes occurring in the simulations 
tend to be transient features without aligned orbits \citep[e.g.][]{Gillet2015}.

The significance of the VPOS can be tested with the help of additional satellite galaxies. In the 
data of the Sloan Digital Sky Survey \citep[SDSS][]{York2000} 15 faint 
and ultra-faint satellite galaxies have been discovered between 2005 and 2010 \citep{Willman2005a,Willman2005b,Belokurov2006,
Belokurov2007,Belokurov2008,Belokurov2009,Belokurov2010,
Sakamoto2006,Zucker2006a,Zucker2006b,Walsh2007,Grillmair2009}.
These additional objects have supported the notion that the majority of MW satellites are part 
of a single halo structure: they independently define a polar plane 
which is closely aligned with the original disk of satellites \citep{Kroupa2010}. However, the VPOS results 
are based on the SDSS survey footprint, which covers the north Galactic pole region and thus the SDSS satellites may be expected 
to lie close to the VPOS. The SDSS survey nevertheless provides additional constraints. 
\citet{Metz2009} noted a deficit of SDSS-discovered satellite galaxies away from the satellite plane
at large galactocentric distances. If the satellites discovered in the SDSS would be drawn from an isotropic distribution, 
the chance of them being oriented in a similarly narrow plane this closely aligned with that defined by the 11 classical 
satellites is small (Pawlowski in prep.). In addition, even though the SDSS survey area was 
extended since data release 7, no new MW satellite galaxies were discovered outside of the VPOS (Pegasus\,III, recently discovered in SDSS DR10 data by \citealt{Kim2015b}, aligns well with the VPOS as will be shown later).

Recently a number of new surveys began to operate, facilitating a wider search for MW satellites beyond the SDSS footprint.
The Pan-STARRS1 survey covers three quarters of the entire sky ($\delta > -30^{\circ}$), 
with much of this area being far away from the VPOS and not covered by SDSS. Two new MW satellite objects
have been discovered in this survey so far: the remote globular cluster PSO~J174.0675-10.8774 \citep{Laevens2014} also known as Crater 
\citep{Belokurov2014}, which happens to lie close to the VPOS \citep{PawlowskiKroupa2014}, and the dwarf satellite 
Triangulum\,II \citep{Laevens2015}.

In recent months, the Stromlo Milky Way Satellite Survey \citep{Jerjen2010}, the Dark Energy Survey \citep[DES;][]{DES2005} and the Survey of the Magellanic Stellar History (SMASH; \citealt{SMASH}; PI D. Nidever) have been revealing a series of new Milky Way companions: two ultra-faint star clusters Kim\,1 \citep{KimJerjen2015} and Kim\,2 \citep{Kim2015}, and two ultra-faint dwarf galaxies: Pegasus\,III \citep{Kim2015b} and Hydra\,II \citep{Martin2015}. Two studies of the DES-Y1A1 survey data announced the discovery of seven \citep[][, hereafter DES15]{Bechtol2015} respectively eight \citep{Koposov2015} new objects, not counting Kim\,2 (or Indus\,I), which has been already found earlier \citep{Kim2015}. Furthermore, \citet{Kim2015c} have discovered an additional object, Horologium II, in the same data set which apparently was overlooked previously. One of the objects discovered in the DES, Eri\,II, is most likely beyond the virial radius of the MW. Fundamental parameters for these new objects are compiled in Table \ref{tab:obsprop}. Fig. \ref{fig:ASP} shows their positions in relation to the MW, the previously known satellites and the Magellanic Stream in an all-sky plot.

Many of the new objects are candidate satellite galaxies although the distinction between star clusters and ultra-faint satellite galaxies based on their size and luminosity alone is becoming increasingly blurry. We therefore refrain from choosing a preferred name, but instead list all suggested names in Table \ref{tab:obsprop}. We will identify the objects by the abbreviations given in the first column, consisting of three letters and a Roman numeral, except Kim\,1 \& 2, which have been unambiguously classified as star clusters.

The alignment of the satellite galaxies in a common structure indicate that they might be dynamically associated, sharing similar orbits. This assumption allows us to predict their PMs \citep{LyndenBell1995,PawlowskiKroupa2013}, measurement of which then provides a crucial test of whether the assumed association is real. The existence of the VPOS and the preferential alignment of streams with this structure indicates such a dynamical association of its constituents. The fact that the 11 classical MW satellites already have PM measurements which show that most of them indeed (co-)orbit in the VPOS further supports this interpretation. The PMs of the new MW satellite objects can thus be predicted in an entirely empirical way \citep{PawlowskiKroupa2013}. This prediction is based solely on the current spatial distribution of the MW satellites and does not require an assumption of a specific MW potential or underlying type of dynamics. Confidence in the method can be drawn from the finding that more precise PM measurements tend to agree better with the predicted PMs \citep{PawlowskiKroupa2013}.

The observed distribution of stellar light (and thus stellar mass) in the newly discovered objects can be used to predict the velocity dispersions of the newly discovered objects. The results depend strongly on which dynamical model (Newtonian dynamics or Modified Newtonian Dynamics, MOND, \citealt{Milgrom1983}) or which dark matter halo scaling relation is assumed \citep{McGaugh2007,Walker2009}. We will provide predictions for all these different cases.

This paper is structured as follows. In Sect. \ref{sect:VPOS} we will compare the positions of the new satellite objects with the VPOS defined by the previously-known MW satellites before measuring the effect of adding the new objects to the VPOS plane fit. Our predictions for the proper motions of the objects are presented in Sect. \ref{sect:propmo}, and the predicted velocity dispersions in Sect. \ref{sect:veldisp}. We end with concluding remarks in Sect. \ref{sect:conclusions}.

\section{The VPOS}
\label{sect:VPOS}

\begin{table*}
\begin{minipage}{180mm}
 \small
 \caption{New plane fit parameters (all objects except confirmed star clusters)}
 \label{tab:planeparams}
 \begin{center}
 \begin{tabular}{@{}lcccccc}
 \hline 
Name & VPOSall & VPOS-3 & VPOS+new & VPOS+new-4 & VPOSsouth & VPOSnorth \\
 \hline
$n \begin{pmatrix} l \\ b \end{pmatrix}$\ [$^{\circ}$] & $\begin{pmatrix}  155.6 \\ -3.3 \end{pmatrix}$  & $\begin{pmatrix}  169.5 \\ -2.8 \end{pmatrix}$  & $\begin{pmatrix}  164.0 \\ -6.9 \end{pmatrix}$  & $\begin{pmatrix}  169.4 \\ -6.1 \end{pmatrix}$  & $\begin{pmatrix}  169.3 \\ -7.2 \end{pmatrix}$  & $\begin{pmatrix}  157.7 \\ -8.9 \end{pmatrix}$\\ 
$D_{\mathrm{MW}}$\ [kpc] & $ 7.9$ & $ 10.4$ & $ 2.5$ & $ 5.4$ & $ 3.6$ & $ 3.0$\\ 
$\Delta$\ [kpc] & $29.3$ & $19.9$ & $30.9$ & $21.3$ & $22.3$ & $36.4$\\ 
$c/a$ & $0.301$ & $0.209$ & $0.313$ & $0.224$ & $0.356$ & $0.475$\\ 
$b/a$ & $0.576$ & $0.536$ & $0.579$ & $0.566$ & $0.619$ & $0.623 $\\ 
$N_{\mathrm{members}}$ & 27 & 24 & 38 & 34 & 19 & 19\\ 
Outliers excluded? & no & yes & no & yes & no & no\\
Includes new objects? & no & no & yes & yes & yes & yes\\
 \hline
 \end{tabular}
 \end{center}
 \small \medskip
Parameters of the plane fits:\\
$\mathbf{n}$: The direction of the normal vector (minor axis) of the best-fit plane in Galactic longitude $l$\ and latitude $b$.\\
$D_{\mathrm{MW}}$: offset of the planes from the center of the MW.\\
$\Delta$: RMS height from the best-fit plane of the MW satellite objects included in the fit.\\
$c/a$\ and $b/a$: short- and intermediate-to-long axis ratios, determined from the RMS heights in the directions of the three axes.\\
$N_{\mathrm{members}}$: Number of objects used for the fits.
\end{minipage}
\end{table*}

If the VPOS is a real structure, then it makes a prediction for the probable locations of new satellites.  Here we compare the positions of the new satellite objects with the VPOS as defined by the previously-known MW satellite galaxies (sect. \ref{sect:VPOSold}). We then determine how the VPOS orientation changes when the new objects are included in the plane fit.

For Kim\,1, Lae\,II, Hor\,II, Hyd\,II, Tri\,II and Peg\,III we will use the positions and distances reported in their respective discovery papers, as listed in Table \ref{tab:obsprop}. For those satellite objects discovered by both DES15 and \citet{Koposov2015} we will use the average of the two distances estimates. The one exception is the star cluster Kim\,2, for which more reliable observational data was obtained in the discovery paper by \citealt{Kim2015}, such that we use their distance measurement.  We found that our results are statistically robust against either using the DES15 or the \citet{Koposov2015} distances.

\subsection{Comparison to the known VPOS}
\label{sect:VPOSold}

As a first step, we have compared the positions of the new MW satellite objects with the VPOS plane orientation given by (i) the 27 previously known MW satellite galaxies (hereafter VPOSall) and (ii)  the fit to 24 of those galaxies, excluding the three outliers Hercules, Ursa Major and Leo I (hereafter VPOS-3). The respective plane parameters from \citealt{Pawlowski2013} are reproduced in the second and third column of Table \ref{tab:planeparams}. As already mentioned by \citet{KimJerjen2015}, \citet{Kim2015}, DES15, \citet{Martin2015} and \citet{Kim2015b}, the new objects lie close to the VPOS. Table \ref{tab:dwarfdist} lists their offsets from the VPOSall and VPOS-3. The average offset of the 13 new objects from the VPOSall plane is only 31\,kpc, the median is 21\,kpc.

Eight objects are particularly close to the best-fit planes. The seven satellite objects Kim\,1, Ret\,II, Hor\,I, Hor\,II, Eri\,III, Pic\,I and Peg\,III have offsets of $\lesssim 20$\,kpc from the VPOSall, and all but Eri II have offsets of only 11\,kpc or less from the VPOS-3. Considering Eri II's large distance of $350\pm30$\,kpc from the MW, it is worth noticing that the two distance ratios $d_{\mathrm{VPOSall}}/d_{\mathrm{MW}}$ and $d_{\mathrm{VPOS-3}}/d_{\mathrm{MW}}$ of less than 14 percent are remarkably small.

The largest offsets are found for the candidate objects Gru I, Hyd II and the star cluster Kim\,2. \citet{Koposov2015}  note that Gru I is located close to a CCD chip gap in the DES survey, such that its parameters are more uncertain than those of the other candidates. A smaller distance from the MW would reduce the offset. Furthermore, in particular the two dwarf galaxy candidates Gru I and Hyd II both have large Galactocentric distances, such that despite their relatively large perpendicular offset from the planes their position vectors are only about $30^{\circ}$\ inclined relative to the planes. In summary, most of the newly reported satellite galaxies and star clusters are qualitatively consistent with the original VPOS.

\subsection{Influence of the DES footprint}
\label{sect:DESfootprint}

\begin{figure}
 \centering
 \includegraphics[width=80mm]{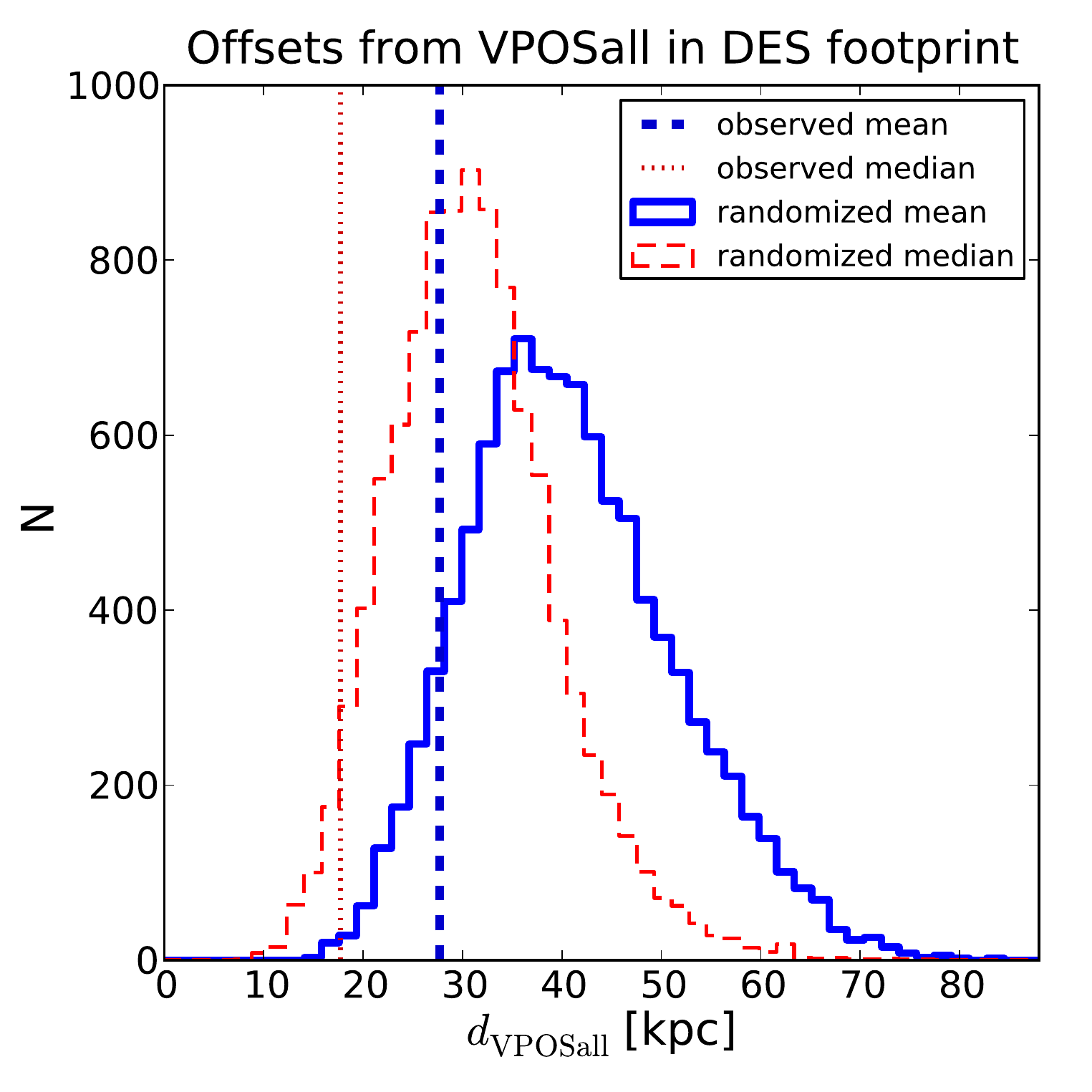}
 \includegraphics[width=80mm]{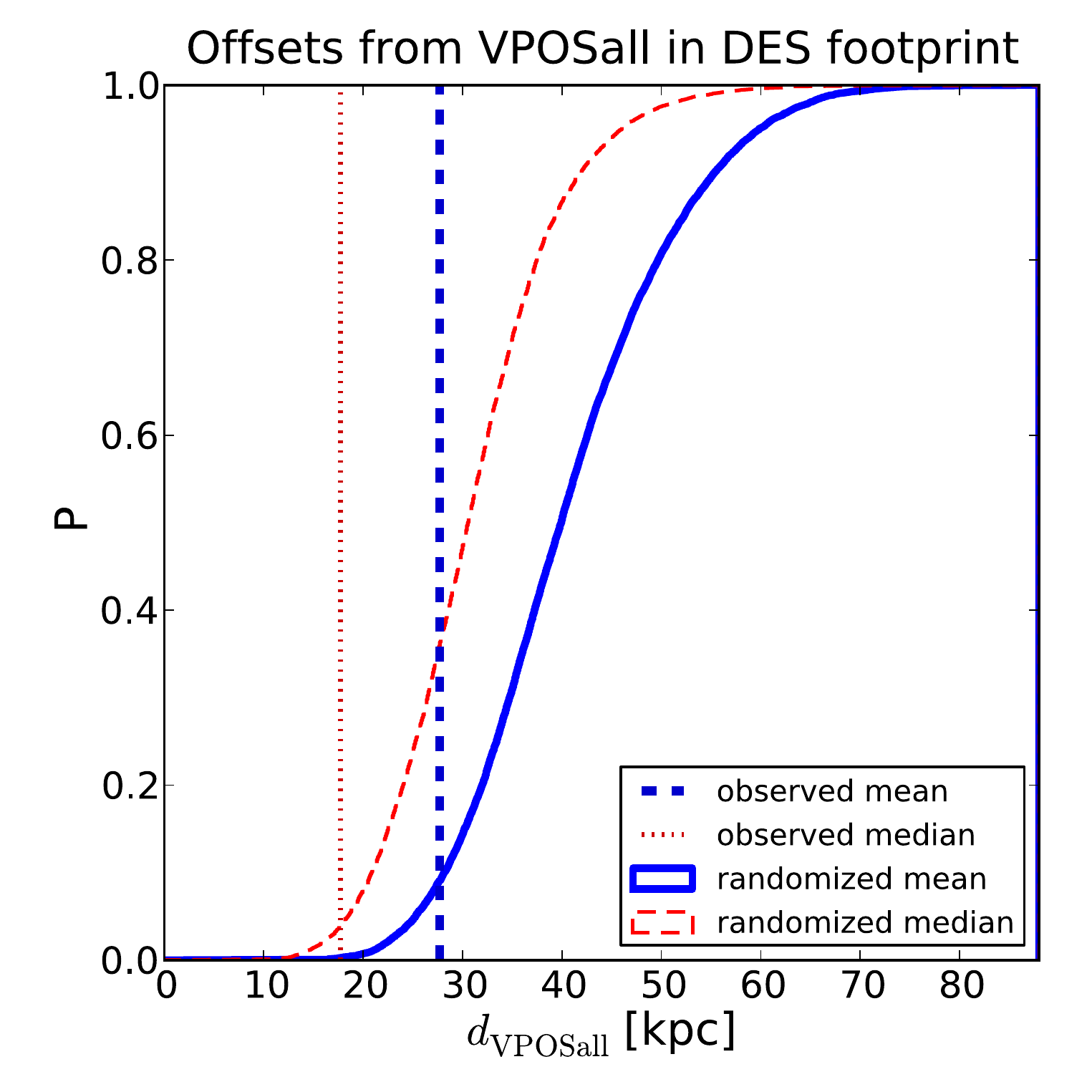}
 \caption{
Histogram (upper panel) and cumulative distribution (lower panel) of mean and median offsets for 10000 randomized distributions of the 10 stellar systems within the DES-Y1A1 footprint, as discussed in Sect. \ref{sect:DESfootprint}. Even though the DES footprint is close to the VPOS, the observed mean and median (thick solid blue and thinner dashed red histograms, respectively) offsets of the 10 objects (vertical dashed and dotted lines, respectively) are smaller than expected for isotropically distributed objects. 
 }
 \label{fig:DESfootprint}
\end{figure}

The region covered by the DES-Y1A1 search area is close to the Magellanic Clouds, and thus not far from the VPOS plane. It might therefore not be surprising that the satellite objects discovered by this survey are close to the VPOS, as argued by DES15. To quantify how much of an alignment is expected due to the survey footprint, and thus to test whether it causes the alignment, we have created 10000 realizations of isotropic DES object positions. To construct these, the angular positions of the ten DES objects have been randomly selected from an isotropic distribution around the Galactic center, while their Galactocentric distances have been preserved. Only positions lying within the region covered by the already observed part of the DES footprint were accepted. Positions outside of this region were randomized again until they were within the survey footprint. This guarantees that each randomized realisation contains ten objects which share the same radial distribution as the observed objects. We have then determined the offsets from the VPOSall plane for each position, and determined the mean and median offsets for each set of ten randomized DES satellite objects.

Figure \ref{fig:DESfootprint} shows the resulting distribution of the mean and the median offsets of these 10000 realisations. The ten observed objects have an average offset of 28\,kpc and a median offset of 18\,kpc. If they were drawn from an isotropic distribution, the expected (average) mean and median would be considerably (about 50\,per cent) larger: 41\,kpc and 31\,kpc, respectively. Of all randomized realisations, 91\,per cent have a larger mean offset, and 96\,per cent have a larger median offset than the observed values. This indicates that the observed alignment is indeed stronger than expected from the current survey footprint alone, but the significance of this conclusion is not extremely high. This might change once the full DES footprint is covered, in particular if at larger distances from the VPOS fewer satellites are discovered.

\subsection{Effects on the best-fit VPOS}

\begin{figure}
 \centering
 \includegraphics[width=80mm]{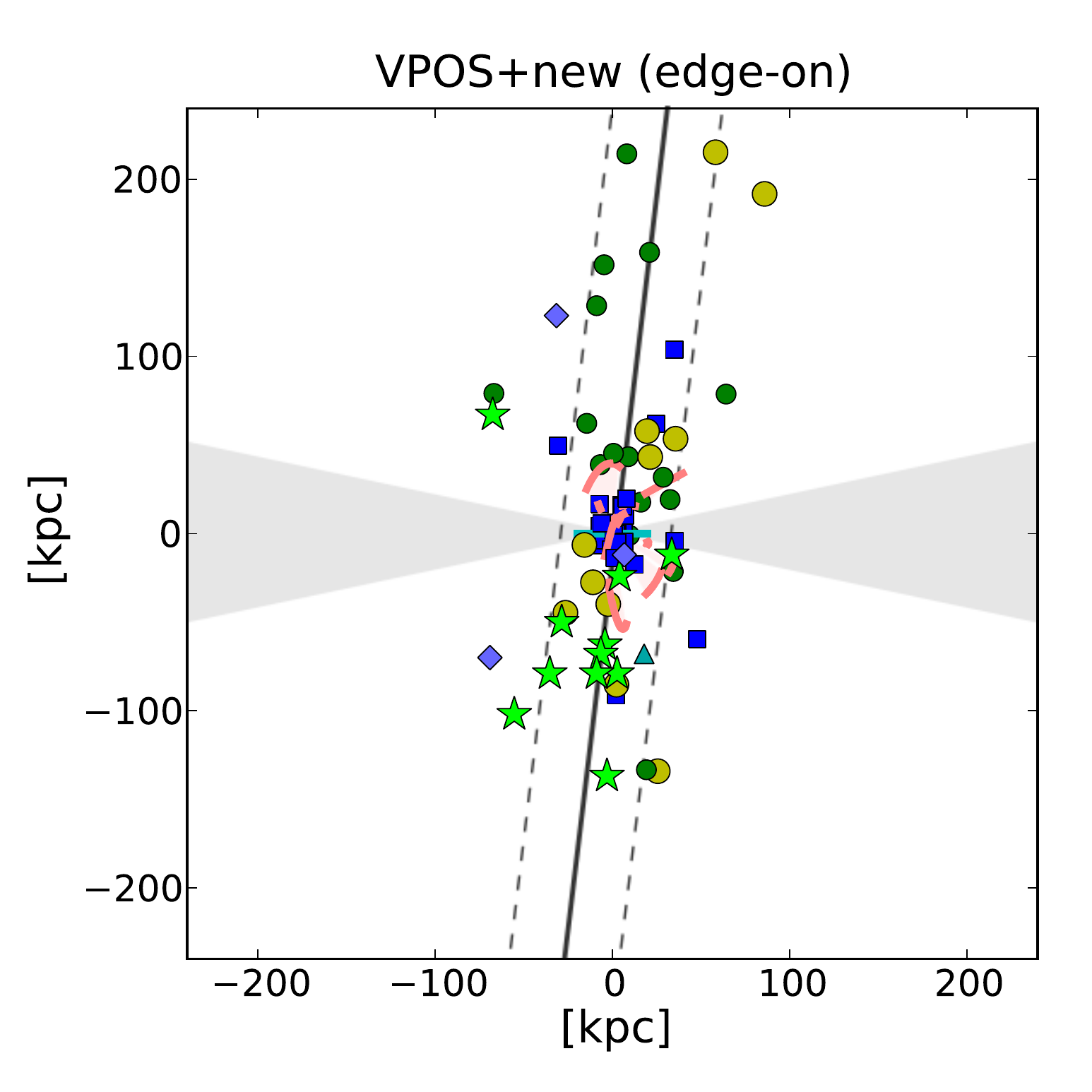}
 \includegraphics[width=80mm]{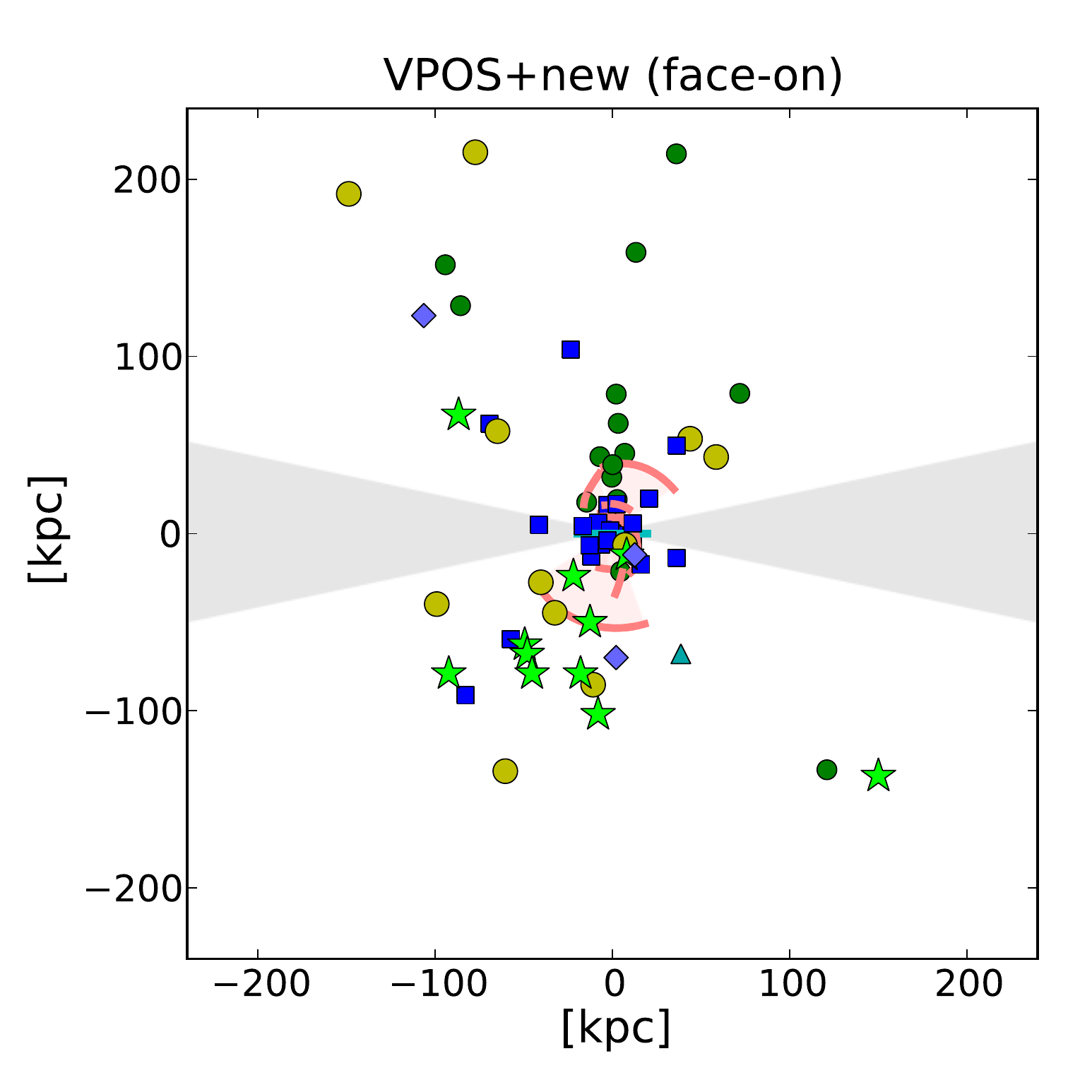}
 \caption{
Distribution of satellite objects in Cartesian coordinates around the MW. The \textit{upper panel} shows the VPOS+new fit edge-on (black solid line, the dashed lines indicate the rms height), the \textit{lower panel} shows a view rotated by $90^{\circ}$, in which the VPOS is oriented approximately face-on. In this view, the MW satellites with measured proper motions preferentially orbit in the clockwise direction. The 11 brightest (classical) MW satellite galaxies are plotted as yellow dots, the fainter satellite galaxies as smaller green dots and globular clusters classified as young halo objects as blue squares. New objects confirmed to be star clusters (PSO~J174.0675-10.8774/Crater in the north, Kim\,1 \& 2 in the south) are plotted as lighter blue diamonds, all other new objects as bright-green stars. The red lines in the center indicate the position and orientation of streams in the MW halo. They preferentially align with the VPOS, but are mostly confined to the innermost regions of the satellite distribution.
Both plots are centred on the MW (cyan line) which is seen edge-on. The grey wedges indicate the region ($\pm12^{\circ}$\ around the MW disk, where satellite galaxies might be obscured by the Galaxy.
}
 \label{fig:VPOS}
\end{figure}

We now update the plane fits by including the new objects. Since we use the same fitting routine and adopt the same parameters as in \citet{Pawlowski2013}, we refer the reader to that paper for further information on the method. 

To be consistent with the previous plane fitting analysis that focussed on satellite galaxies, in the following we add only those 11 new objects to the plane fit sample which are likely but still unconfirmed satellite galaxies. The two objects Kim\,1 \& 2, which have already been identified as star clusters through follow-up observations are not included. We also exclude Eri\,II  from the fit because its large distance estimate places it outside of the virial radius of the MW ($\approx250$\,kpc), thus should not be considered a satellite galaxy of the Milky Way.

For the full VPOS sample (called VPOS+new in Table \ref{tab:planeparams}) we use all 27 previously identified satellites plus the 11 new satellite objects. Our analysis shows that even though the sample size used for the fit has increased by 40\,percent, and the number of objects in the southern Galactic hemisphere doubled, the VPOS parameters remain essentially unchanged. The direction of the plane normal $n$\ changes by only about $9^{\circ}$, to larger Galactic longitude and slightly lower Galactic latitude. This normal is closely aligned (within $9^{\circ})$\ with the normal direction of the `classical' disk of satellites defined by the 11 brightest MW satellites, which points to $(l, b) = (157.3^{\circ}, -12.7^{\circ})$. Because these are the brightest MW satellites, their distribution should be the least affected by biases due to uneven sky coverages, but low-number statistics are a concern.

Our plane fit routine does not require the fitted plane to pass through the MW center. This extra freedom acts as a consistency check for whether a found plane can be dynamically stable\footnote{If a satellite plane has a large offset from its host's center (such as Plane 2 in \citealt{ShayaTully2013}), the satellites can not orbit within the plane and the arrangement must be a transient feature.}. It is therefore arguably the most interesting finding that the new offset from the MW center $D_{\mathrm{MW}}$, is reduced from 7.9 to only 2.5\,kpc. This might indicate that it needed more satellites in the southern Galactic hemisphere to `balance' VPOS out in the MW center. Despite the overall shift to smaller MW offsets, both measures of the plane thickness, the root-mean-square height $\Delta$\ and the short-to-long axis ratio $c/a$, remain almost unchanged (they both increase by a few per cent) if the additional satellite galaxy candidates are included in the fit. The resulting values $\Delta = 30.9$\,kpc and $c/a = 0.313$\ are both at least 50\,per cent smaller than those expected for the same number of satellites objects drawn from isotropic distributions or from sub-halo distributions in cosmological simulations of MW equivalents \citep[see e.g. fig. 3 of][]{PawlowskiMcGaugh2014b}.

The distances of the satellites from this new best-fit plane are also compiled in Table \ref{tab:dwarfdist}. Not surprisingly, the new objects have smaller offsets from the VPOS plane if they are included in the plane fit. Since the plane orientation has not changed dramatically compared to the VPOSall, the same objects tend to be closest to the new fit, too.

In addition to fitting a plane to all 27 known plus the 11 new satellite candidates, we have also constructed a sample analogous to the VPOS-3 by excluding the four outliers with offsets of more than 50\,kpc from the VPOS+New fit. These are the three previously-identified outliers Hercules, Ursa Major I, and Leo I, as well as the newly discovered object Hyd II. We note that Leo\,I has a proper motion indicating that it does not orbit within the VPOS. It might not even be a true MW satellite given its large Galactocentric distance and high velocity. This VPOS+New-4 has a very similar orientation as the VPOS+New (inclined by about $6^{\circ}$) but is significantly thinner ($\Delta = 21.3$\,kpc, $c/a = 0.224$). It is almost identical in orientation and thickness to the VPOS-3, but also has a smaller offset from the MW center.

As reported before \citep{Pawlowski2013}, the VPOS plane is oriented almost exactly like the current orbital plane of the LMC, which has an orbital pole (direction of angular momentum) pointing towards $(l, b) = (175.4^{\circ}, -5.7^{\circ})$\ \citep{PawlowskiKroupa2013}, less than $12^{\circ}$\ inclined with respect to the VPOS+new and only $6^{\circ}$\ with respect to the VPOS+New-4. This supports the almost 40 years old notion by \citet{KunkelDemers1976} and \citet{LyndenBell1976} that most MW satellites seem to share the same orbital plane as the Magellanic Clouds.

Fig. \ref{fig:VPOS} shows the distribution of known satellite objects around the MW from both an edge-on and face-on views of the updated VPOS+new. Newly discovered satellite objects are plotted as light blue diamonds if they have been confirmed to be star clusters, or as bright green stars. It is obvious that most of the recently discovered satellite galaxy candidates lie within the VPOS plane at intermediate distances from the MW. The majority populate a region where previously-known classical MW satellites (yellow dots in the figure) are situated\footnote{The LMC and SMC, Carina, Fornax and Sculptor}. This is illustrated by the face-on VPOS+New view in the lower panel of Fig. \ref{fig:VPOS}, in which the majority of new discoveries populate the lower (southern) left quadrant of the VPOS plane. 

This panel also illustrates how isolated the new MW satellite Pegasus III is. As pointed out by \citet{Kim2015b}, its only known close neighbour at a distance of about 37\,kpc is Pisces II. The next-closest known MW satellites are found at distances of 170\,kpc or more. The whole lower (southern) right quadrant of the VPOS in the lower panel of Fig. \ref{fig:VPOS} appears to be surprisingly devoid of known MW satellites. None of the 11 classical MW satellites falls into this region at Galactocentric distances beyond about 30\,kpc (the closest one, Sagittarius, lies at the edge of this quadrant but is known to orbit perpendicular to the VPOS). Interestingly, part of this quadrant was surveyed by the SEGUE extension of the SDSS survey, within which Pisces II, the only previously known faint MW satellite galaxies in this region, was found. The next-closest known structure, also found in the data of the SDSS survey, is the Pisces overdensity \citep{Watkins2009} at a distance of 130\,kpc from Pegasus III (green triangle in Fig. \ref{fig:VPOS}). The overdensity might be the remnant of a dwarf galaxy or star cluster currently being disrupted, and is also aligned with the Magellanic Stream and the VPOS.

\subsection{Dividing the VPOS: north-south differences?}

For a long time, the majority of known MW satellites galaxies lay in the Galactic north because the SDSS initially concentrated on that part of the sky. With the discovery of new satellite galaxy candidates, the numbers of southern objects approaches that of the northern ones (19 each, if all of the new candidate objects turn out to be bona-fide satellite galaxies). The comparable sample size allows us to fit the VPOS for the southern and northern satellite populations separately. Such a split might reveal whether the VPOS is systematically tilted or bent, for example due to precession of the satellites on their orbits or because of non-symmetric infall of the VPOS-satellites. Due to the Galactic disc which possibly obscures satellites along the Galactic equator, splitting the satellites into a northern and southern sample is the most sensible separation.

The resulting plane-fit parameters for the two sub-samples are similar to each other and to the VPOS+New plane fitted to all objects (see Table \ref{tab:planeparams}). Both the northern and the southern VPOS plane fits are inclined by only $6^{\circ}$\ relative to the total VPOS+New plane fit. They both are almost polar (the normal vectors have Galactic latitudes of $b = -7^{\circ}$\ and $-9^{\circ}$, for the southern and the northern satellites, respectively), have a very similar orientation (inclined by less than $12^{\circ}$), have essentially the same small offset from the Galactic center ($D_{\mathrm{MW}} \approx 3$\,kpc) and are similarly thin ($\Delta_{\mathrm{rms}} = 22$\ versus 36\,kpc). However, while the thinner, southern plane shares its orientation with the orbital plane of the LMC (and the VPOS-3 plane fits) to within $6^{\circ}$, the northern one is more inclined ($18^{\circ}$).

Interestingly, the orientation and thickness of the northern plane is strongly affected by the four outliers, which are all northern objects. If they are excluded from the northern VPOS fit, the resulting plane parameters are extremely similar to those of the southern fit: the normal to the best-fit plane points to $(l, b) = (168.5^{\circ}, -3.3^{\circ})$, the root-mean-square height of the plane is $\Delta_{\mathrm{rms}} = 20$\,kpc, but the offset from the MW center is $D_{\mathrm{MW}} = 11$\,kpc. Even without excluding the outliers from the northern satellite sample, the plane parameters of the northern (VPOSnorth) and the southern (VPOSsouth) fits agree well. We therefore conclude that the current data suggest that the northern and the southern part of the VPOS have similar properties and orientations. This might also indicate that observational biases -- which are different for the northern and southern hemispheres -- do not affect the determined plane orientations significantly.

\begin{table*}
\begin{minipage}{180mm}
 \caption{Galactocentric distances, Cartesian positions and offsets of MW satellites from the VPOS plane fits (in kpc)}
 \label{tab:dwarfdist}
 \begin{center}
 \begin{tabular}{@{}lcccccccccccc}
 \hline 
Name & $d_{\mathrm{MW}}$ & $x$ &  $y$ &  $z$ & $d_{\mathrm{VPOSall}}$ & $d_{\mathrm{VPOS-3}}$ & $d_{\mathrm{VPOS+New}}$ & $d_{\mathrm{VPOS+New-4}}$ & $d_{\mathrm{VPOSsouth}}$ & $d_{\mathrm{VPOSnorth}}$\\
 \hline
The Galaxy &  0.0 &  0.0 &  0.0 &  0.0 & $       8 $ & $      10 $ & $       3 $ & $       5 $ & $       4 $ & $       3 $ \\ 
Canis Major & 13.4 & -11.9 & -6.2 & -1.0 & $ \bf    1 $ & $ \bf    1 $ & $ \bf    7 $ & $ \bf    5 $ & $ \bf    7 $ & $      11 $ \\ 
Sagittarius dSph & 18.4 & 17.1 &  2.5 & -6.4 & $ \bf   23 $ & $ \bf   28 $ & $ \bf   19 $ & $ \bf   22 $ & $ \bf   21 $ & $      12 $ \\ 
Segue (I) & 27.9 & -19.4 & -9.5 & 17.7 & $ \bf    5 $ & $ \bf    7 $ & $ \bf   12 $ & $ \bf   11 $ & $      12 $ & $ \bf   15 $ \\ 
Ursa Major II & 38.0 & -30.6 & 11.6 & 19.2 & $ \bf   26 $ & $ \bf   23 $ & $ \bf   30 $ & $ \bf   27 $ & $      29 $ & $ \bf   35 $ \\ 
Bootes II & 39.5 &  6.6 & -1.7 & 38.9 & $ \bf   17 $ & $ \bf   19 $ & $ \bf   14 $ & $ \bf   17 $ & $      14 $ & $ \bf   10 $ \\ 
Segue II & 40.8 & -31.8 & 13.9 & -21.4 & $ \bf   29 $ & $ \bf   26 $ & $ \bf   35 $ & $ \bf   32 $ & $ \bf   33 $ & $      41 $ \\ 
Willman 1 & 42.9 & -27.7 &  7.6 & 31.8 & $ \bf   22 $ & $ \bf   20 $ & $ \bf   25 $ & $ \bf   22 $ & $      25 $ & $ \bf   29 $ \\ 
Coma Berenices & 44.9 & -10.6 & -4.3 & 43.4 & $ \bf    3 $ & $ \bf    3 $ & $ \bf    1 $ & $ \bf    1 $ & $       2 $ & $ \bf    4 $ \\ 
Bootes III & 45.8 &  1.3 &  6.9 & 45.3 & $ \bf    9 $ & $ \bf   13 $ & $ \bf    7 $ & $ \bf   11 $ & $       8 $ & $ \bf    2 $ \\ 
LMC & 50.0 & -0.6 & -41.8 & -27.5 & $ \bf   24 $ & $ \bf   16 $ & $ \bf   11 $ & $ \bf   10 $ & $ \bf    9 $ & $      10 $ \\ 
SMC & 61.2 & 16.5 & -38.5 & -44.7 & $ \bf   38 $ & $ \bf   33 $ & $ \bf   25 $ & $ \bf   25 $ & $ \bf   24 $ & $      23 $ \\ 
Bootes (I) & 64.0 & 14.8 & -0.8 & 62.2 & $ \bf   26 $ & $ \bf   29 $ & $ \bf   25 $ & $ \bf   27 $ & $      25 $ & $ \bf   20 $ \\ 
Draco & 75.9 & -4.4 & 62.2 & 43.2 & $ \bf   21 $ & $ \bf    4 $ & $ \bf   15 $ & $ \bf    6 $ & $       9 $ & $ \bf   27 $ \\ 
Ursa Minor & 77.8 & -22.2 & 52.0 & 53.5 & $ \bf   33 $ & $ \bf   19 $ & $ \bf   28 $ & $ \bf   21 $ & $      24 $ & $ \bf   38 $ \\ 
Sculptor & 86.0 & -5.2 & -9.8 & -85.3 & $ \bf    2 $ & $ \bf    3 $ & $ \bf   10 $ & $ \bf    7 $ & $ \bf    8 $ & $      16 $ \\ 
Sextans (I) & 89.0 & -36.7 & -56.9 & 57.8 & $ \bf    2 $ & $ \bf   13 $ & $ \bf    9 $ & $ \bf   14 $ & $      16 $ & $ \bf    5 $ \\ 
Ursa Major (I) & 101.6 & -61.1 & 19.8 & 78.7 & $ \bf   53 $ & $      51 $ & $ \bf   54 $ & $      51 $ & $      54 $ & $ \bf   57 $ \\ 
Carina & 106.8 & -25.1 & -95.9 & -39.8 & $ \bf   25 $ & $ \bf    1 $ & $ \bf    2 $ & $ \bf    6 $ & $ \bf    6 $ & $       9 $ \\ 
Hercules & 126.1 & 84.1 & 50.7 & 79.1 & $ \bf   71 $ & $      94 $ & $ \bf   83 $ & $      93 $ & $      89 $ & $ \bf   68 $ \\ 
Fornax & 149.3 & -41.3 & -51.0 & -134.1 & $ \bf   17 $ & $ \bf   30 $ & $ \bf   40 $ & $ \bf   42 $ & $ \bf   41 $ & $      40 $ \\ 
Leo IV & 154.8 & -15.1 & -84.8 & 128.6 & $ \bf   39 $ & $ \bf   18 $ & $ \bf   29 $ & $ \bf   21 $ & $      18 $ & $ \bf   37 $ \\ 
Canes Venatici II & 160.6 & -16.5 & 18.6 & 158.7 & $ \bf    6 $ & $ \bf    2 $ & $ \bf    1 $ & $ \bf    3 $ & $       5 $ & $ \bf    4 $ \\ 
Leo V & 178.6 & -21.5 & -91.9 & 151.7 & $ \bf   37 $ & $ \bf   14 $ & $ \bf   28 $ & $ \bf   18 $ & $      16 $ & $ \bf   38 $ \\ 
Pisces II & 181.1 & 14.9 & 121.7 & -133.3 & $ \bf   38 $ & $ \bf    4 $ & $ \bf   34 $ & $ \bf   17 $ & $ \bf   19 $ & $      57 $ \\ 
Canes Venatici (I) & 217.5 &  2.1 & 37.0 & 214.3 & $ \bf    6 $ & $ \bf   17 $ & $ \bf   20 $ & $ \bf   24 $ & $      19 $ & $ \bf   14 $ \\ 
Leo II & 235.9 & -77.3 & -58.3 & 215.2 & $ \bf   26 $ & $ \bf   46 $ & $ \bf   30 $ & $ \bf   38 $ & $      42 $ & $ \bf   20 $ \\ 
Leo I & 257.4 & -123.6 & -119.3 & 191.7 & $ \bf   45 $ & $      83 $ & $ \bf   60 $ & $      76 $ & $      79 $ & $ \bf   41 $ \\ \hline
Kim I & 19.1 & -2.7 & 14.4 & -12.3 & $       2 $ & $       4 $ & $       6 $ & $       1 $ & $       3 $ & $      13 $ \\ 
Ret II & 33.0 & -9.7 & -20.4 & -24.1 & $       6 $ & $       3 $ & $ \bf    4 $ & $ \bf    3 $ & $ \bf    4 $ & $       7 $ \\ 
Lae 2 (Tri II) & 36.6 & -29.8 & 17.4 & -12.2 & $      27 $ & $      23 $ & $ \bf   32 $ & $ \bf   28 $ & $ \bf   30 $ & $      39 $ \\ 
Tuc II & 59.2 & 24.4 & -20.4 & -49.9 & $      36 $ & $      35 $ & $ \bf   26 $ & $ \bf   28 $ & $ \bf   27 $ & $      20 $ \\ 
Hor II & 80.0 & -9.6 & -48.7 & -62.7 & $      16 $ & $       7 $ & $ \bf    1 $ & $ \bf    1 $ & $ \bf    2 $ & $       4 $ \\ 
Hor I & 83.5 & -7.2 & -48.0 & -67.9 & $      18 $ & $       9 $ & $ \bf    2 $ & $ \bf    0 $ & $ \bf    1 $ & $       3 $ \\ 
Phe II & 88.1 & 28.7 & -27.2 & -78.8 & $      41 $ & $      40 $ & $ \bf   28 $ & $ \bf   30 $ & $ \bf   29 $ & $      23 $ \\ 
Eri III & 91.2 & -4.3 & -46.0 & -78.7 & $      19 $ & $      11 $ & $ \bf    3 $ & $ \bf    1 $ & $ \bf    1 $ & $       4 $ \\ 
Kim II (Ind I) & 98.9 & 67.5 & -17.3 & -70.2 & $      72 $ & $      76 $ & $      63 $ & $      67 $ & $      66 $ & $      55 $ \\ 
Grus 1 & 116.4 & 50.6 & -23.0 & -102.2 & $      57 $ & $      59 $ & $ \bf   45 $ & $ \bf   48 $ & $ \bf   48 $ & $      38 $ \\ 
Pic I & 121.9 & -28.1 & -88.2 & -79.2 & $      15 $ & $       5 $ & $ \bf    8 $ & $ \bf   14 $ & $ \bf   14 $ & $       6 $ \\ 
Hydra II & 129.0 & 40.8 & -102.4 & 66.8 & $      92 $ & $      72 $ & $ \bf   79 $ & $      71 $ & $      70 $ & $ \bf   85 $ \\ 
Pegasus III & 203.1 & 44.3 & 143.5 & -136.8 & $      21 $ & $      21 $ & $ \bf   13 $ & $ \bf    8 $ & $ \bf    6 $ & $      40 $ \\ 
Eri II & 365.0 & -86.2 & -211.4 & -284.7 & $       4 $ & $      50 $ & $      51 $ & $      69 $ & $      66 $ & $      35 $ \\ 
 \hline
 \end{tabular}
 \end{center}
 \small \medskip
MW satellite distances from the MW ($d_{\mathrm{MW}}$), Galactocentric Cartesian $x$, $y$, and $z$\ positions and offsets from the different best-fit VPOS planes: all 27 satellite galaxies considered in \citet{Pawlowski2013} (VPOSall), the same sample but excluding three outliers (VPOS-3) and the fit to the 27 known and confirmed MW satellite galaxies plus all new objects except those which have been identified as star clusters (VPOS+new, i.e. the VPOSall sample plus Ret\,II, Lae\,2, Tuc\,II, Hor\,I and II, Phe\,II, Eri\,III, Gru\,I, Pic\,I, Hyd\,II and Peg\,III). For those of the new objects discovered by both B15 and K15 we assume that their Heliocentric distances are the average of the two distance estimates when fitting the plane. If an object was included in the plane fit its respective offset if printed in boldface.
\end{minipage}
\end{table*}

\subsection{The distant object Eri II}

The two current distance estimates for Eri II are 330 (DES15) or 380\,kpc \citep{Koposov2015}. These place the object barely outside of the virial radius expected for the MW (about 250 to 300\,kpc). If its position just beyond the MW halo is confirmed, this would make Eri II the closest known non-satellite dwarf galaxy in the Local Group, closer than the neighbours Phoenix and Leo T at about 420\,kpc distance. 

Eri II aligns well with the VPOS plane, even though it was excluded from the fit due to its large distance. In this regard it might be interesting to check where Eri II lies with respect to the dominating plane of non-satellite galaxies in the LG, termed LG plane 1 \citep[LGP1, ][]{Pawlowski2013,PawlowskiMcGaugh2014a}. LGP1 consists of about 10 LG dwarf galaxies which have distances of more than 300\,kpc from both the MW and M31. It is an extremely narrow plane, with root-mean-square height of about 50\,kpc and a maximum diameter of 2\,Mpc, which stretches from M31 to the MW and beyond. Approximately in Eri II's direction LGP1 has its closest approach to the MW, of about 180\,kpc. Depending on its exact position, Eri II has an offset of about 100\,kpc from LGP1. It therefore may potentially be a member of this structure. Assuming Eri II to be part of and perfectly aligned with LGP1, we would have predicted it to be closer than the current estimates by about 50 to 100\,kpc. Measuring Eri II's line-of-sight velocity might shed more light on whether an association with LGP1 is likely. All LGP1 dwarf galaxies have line-of-sight velocities relative to the local standard of rest which follow the Magellanic Stream velocity at their positions (see Fig. \ref{fig:MagStream}). If Eri II follows the same trend, its velocity should be in the range of 50 to $250\,\mathrm{km\,s}^{-1}$.

The spatial position of Eri II thus places it at the intersection of LGP1 and the VPOS, a very interesting position holding the promise that Eri II might provide important clues on if and how the VPOS is connected to the larger-scale structure of dwarf galaxies in the LG.

\section{Predicted proper motions}
\label{sect:propmo}

\begin{table*}
\begin{minipage}{180mm}
 \caption{Predicted PMs of MW satellite objects assuming they orbit within the VPOS-3}
 \label{tab:propmo}
 \begin{center}
 \begin{tabular}{@{}lcccccccc}
  \hline
  Name & $l$ & $b$ & $r_{\sun}$ & $\min(v_{\mathrm{los}})$ & $\theta_{\mathrm{VPOS-3}}^{\mathrm{predicted}}$ & $[v_{\mathrm{min}},~v_{\mathrm{max}}]$ & 
 $\begin{pmatrix} \mu_{\alpha} \cos \delta \\ \mu_{\delta} \end{pmatrix}_{\mathrm{co}}$ & 
 $\begin{pmatrix} \mu_{\alpha} \cos \delta \\ \mu_{\delta} \end{pmatrix}_{\mathrm{counter}}$ 
 \\
   & $[^\circ]$ & $[^\circ]$ & [kpc] & $[\mathrm{km\,s}^{-1}]$ & $[^\circ]$ & $[\mathrm{km\,s}^{-1}]$ & 
 $[\mathrm{mas\,yr}^{-1}]$ & 
 $[\mathrm{mas\,yr}^{-1}]$ \\
 \hline
Kim 1 &  68.5 & -38.4 &  20 & $ -182 $ & $ 17.8 $ &  [50,~595] & $\begin{pmatrix} [+0.92,~+4.55] \\ [-2.17,~-6.72] \end{pmatrix}$ & $\begin{pmatrix} [+0.26,~-3.36] \\ [-1.34,~+3.21] \end{pmatrix}$\\
Ret II &  266.3 & -49.7 &  32 & $ 168 $ & $ 11.9 $ &  [50,~518] & $\begin{pmatrix} [+1.43,~+4.41] \\ [-0.66,~-1.61] \end{pmatrix}$ & $\begin{pmatrix} [+0.80,~-2.18] \\ [-0.46,~+0.48] \end{pmatrix}$\\
Tuc II &  328.1 & -52.3 &  63 & $  81 $ & $ 25.3 $ &  [50,~455] & $\begin{pmatrix} [+0.33,~+0.74] \\ [-0.90,~-2.20] \end{pmatrix}$ & $\begin{pmatrix} [+0.23,~-0.18] \\ [-0.59,~+0.71] \end{pmatrix}$\\
Hor II &  268.5 & -52.2 &  79 & $ 160 $ & $ 2.4 $  &  [50,~429] & $\begin{pmatrix} [+0.57,~+1.50] \\ [-0.31,~-0.68] \end{pmatrix}$ & $\begin{pmatrix} [+0.32,~-0.61] \\ [-0.21,~+0.17] \end{pmatrix}$\\
Hor I &  271.4 & -54.7 &  83 & $ 151 $ & $ 1.0 $ &  [50,~421] & $\begin{pmatrix} [+0.54,~+1.39] \\ [-0.33,~-0.74] \end{pmatrix}$ & $\begin{pmatrix} [+0.32,~-0.53] \\ [-0.22,~+0.19] \end{pmatrix}$\\
Phe II &  323.7 & -59.7 &  89 & $  77 $ & $ 19.5 $ &  [50,~423] & $\begin{pmatrix} [+0.31,~+0.73] \\ [-0.59,~-1.35] \end{pmatrix}$ & $\begin{pmatrix} [+0.20,~-0.21] \\ [-0.39,~+0.37] \end{pmatrix}$\\
Eri III &  275.0 & -59.6 &  91 & $ 132 $ & $ 0.3 $ &  [50,~409] & $\begin{pmatrix} [+0.49,~+1.19] \\ [-0.37,~-0.81] \end{pmatrix}$ & $\begin{pmatrix} [+0.29,~-0.41] \\ [-0.24,~+0.20] \end{pmatrix}$\\
Kim 2 &  347.2 & -42.1 & 105 & $  38 $ & $ 41.9 $ & [50,~390] & $\begin{pmatrix} [+0.03,~+0.08] \\ [-0.60,~-1.28] \end{pmatrix}$ & $\begin{pmatrix} [+0.01,~-0.04] \\ [-0.40,~+0.29] \end{pmatrix}$\\
Gru I &  338.7 & -58.2 & 120 & $  49 $ & $ 24.9 $ &  [50,~369] & $\begin{pmatrix} [+0.19,~+0.41] \\ [-0.49,~-1.00] \end{pmatrix}$ & $\begin{pmatrix} [+0.12,~-0.10] \\ [-0.32,~+0.19] \end{pmatrix}$\\
Hyd II &  295.6 & 30.5 & 132 & $ 187 $ & $ 28.9 $ &  [50,~361] & $\begin{pmatrix} [-0.18,~+0.03] \\ [-0.09,~+0.36] \end{pmatrix}$ & $\begin{pmatrix} [-0.25,~-0.46] \\ [-0.24,~-0.69] \end{pmatrix}$\\
Pic I &  257.4 & -41.2 & 120 & $ 191 $ & $ 7.1 $ &  [50,~361] & $\begin{pmatrix} [+0.35,~+0.90] \\ [-0.11,~-0.16] \end{pmatrix}$ & $\begin{pmatrix} [+0.18,~-0.36] \\ [-0.10,~-0.05] \end{pmatrix}$\\
Peg III &  69.8 & -41.8 & 205 & $ -174 $ & $ 2.9 $ &  [50,~273] & $\begin{pmatrix} [+0.10,~+0.24] \\ [-0.22,~-0.40] \end{pmatrix}$ & $\begin{pmatrix} [+0.04,~-0.10] \\ [-0.13,~+0.05] \end{pmatrix}$\\
Eri II &  249.8 & -51.6 & 363 & $ 154 $ & $ 9.3 $ & [50,~105] & $\begin{pmatrix} [+0.12,~+0.15] \\ [-0.07,~-0.08] \end{pmatrix}$ & $\begin{pmatrix} [+0.07,~+0.04] \\ [-0.06,~-0.05] \end{pmatrix}$\\
  \hline
 \end{tabular}
 \end{center}
 \small \smallskip
$l$, $b$, $r_{\sun}$: Heliocentric position of the MW satellite object in Galactic longitude and latitude and radius.\\
$\min(v_{\mathrm{los}})$: heliocentric line-of-sight velocity that minimizes the line-of-sight component of its Galactocentric velocity (i.e. the negative of the Galactocentric solar velocity component in the direction of the object).\\
$\theta_{\mathrm{VPOS-3}}^{\mathrm{predicted}}$: The smallest-possible inclination between the orbital plane of the satellite galaxy and the VPOS-3 plane, defined by the angle between the Galactocentric position and the plane normal.\\
$v_{\mathrm{min}}$, $v_{\mathrm{max}}$: the range of PMs is constrained by adopting this range of minimum and maximum absolute speeds for the objects. The minimum is set to $50\,\mathrm{km\,s}^{-1}$, the maximum determined by requiring the object to be approximately bound to the MW (see \citealt{PawlowskiKroupa2013} for details).\\
$\begin{pmatrix} \mu_{\alpha} \cos \delta \\ \mu_{\delta} \end{pmatrix}_{\mathrm{co}}$\ and $\begin{pmatrix} \mu_{\alpha} \cos \delta \\ \mu_{\delta} \end{pmatrix}_{\mathrm{counter}}$: predicted PM range if co- or counter-orbiting.\\
\end{minipage}
\end{table*}

\begin{figure*}
 \centering
\includegraphics[width=58mm]{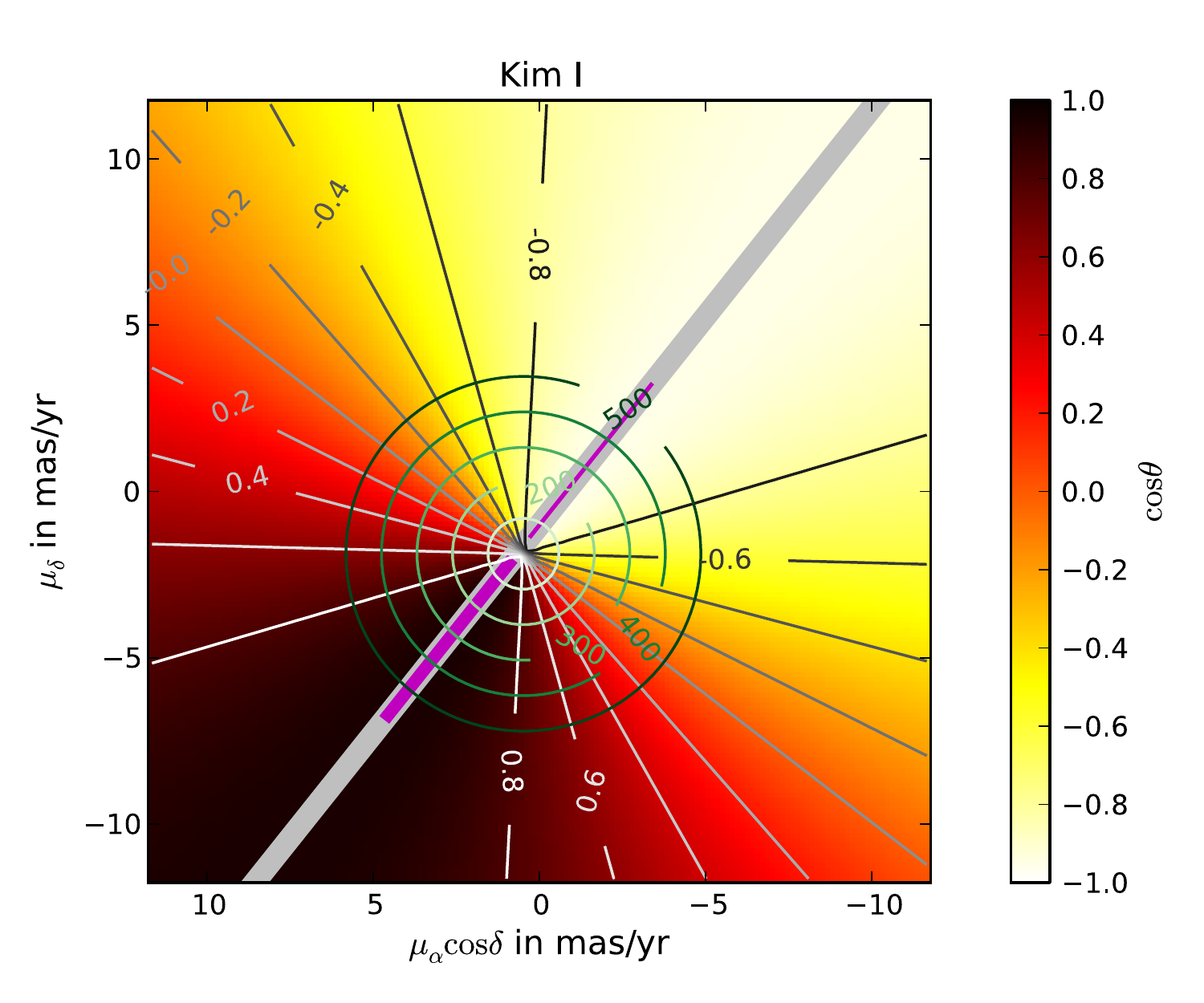}
\includegraphics[width=58mm]{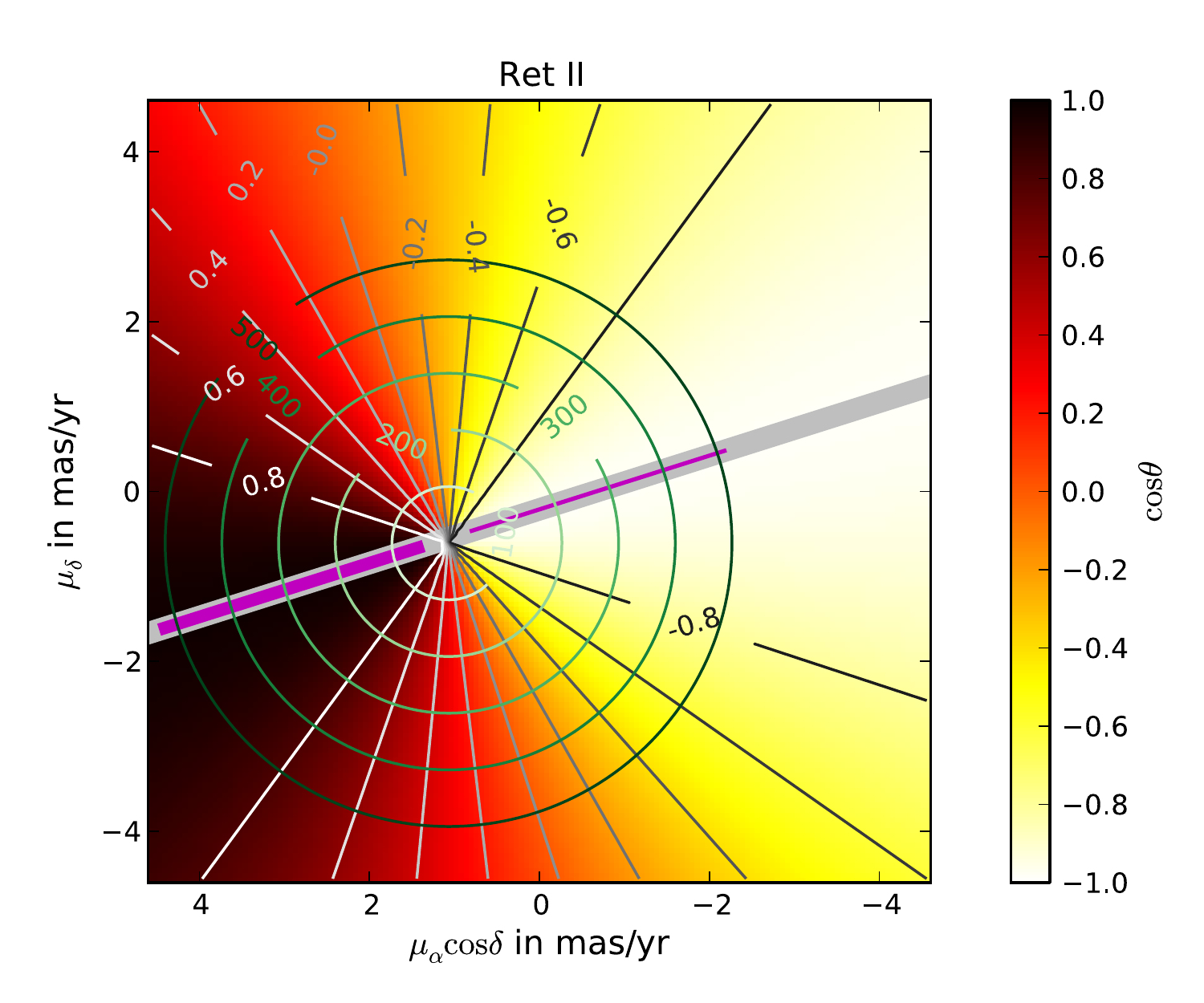}
\includegraphics[width=58mm]{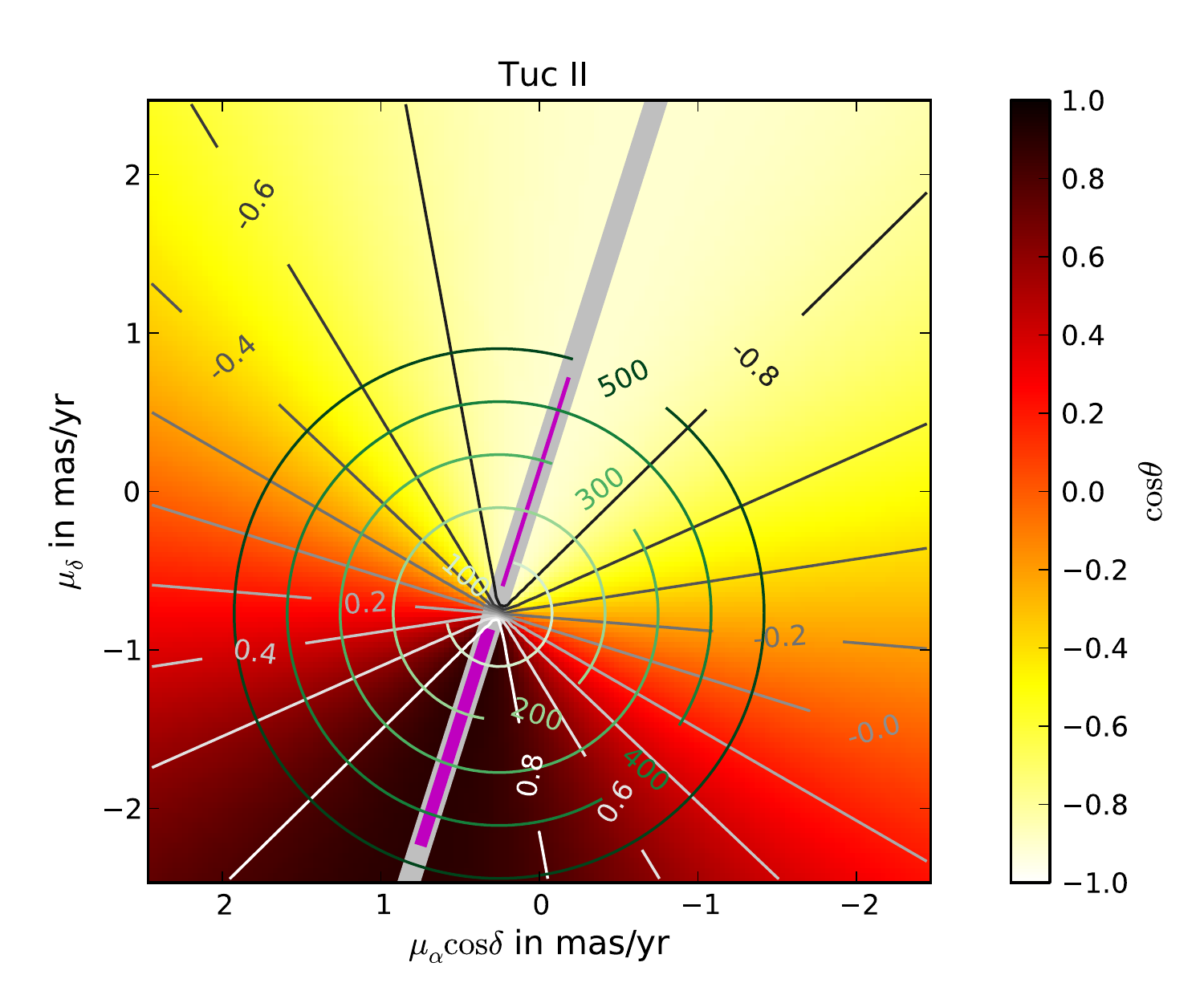}
\includegraphics[width=58mm]{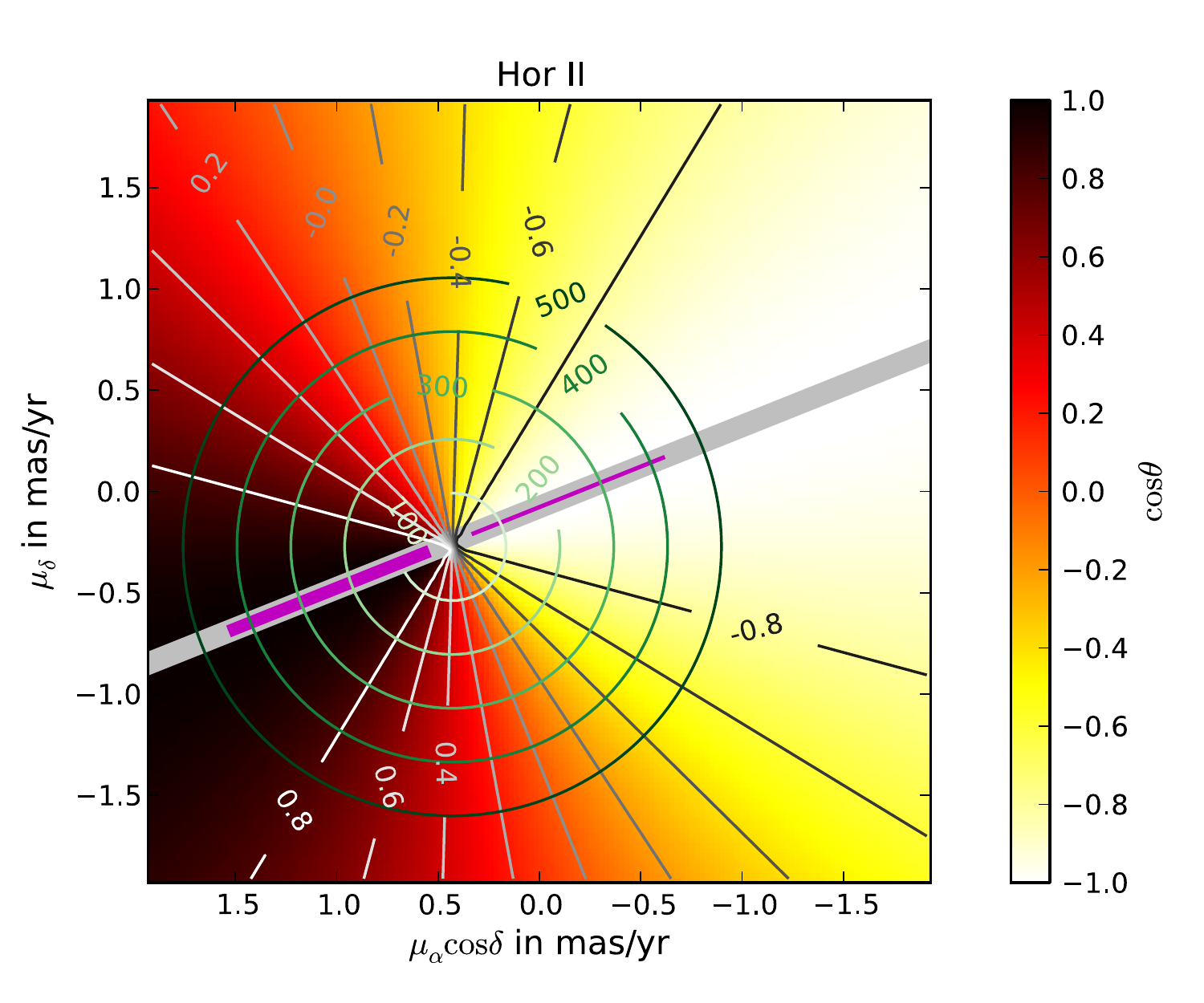}
\includegraphics[width=58mm]{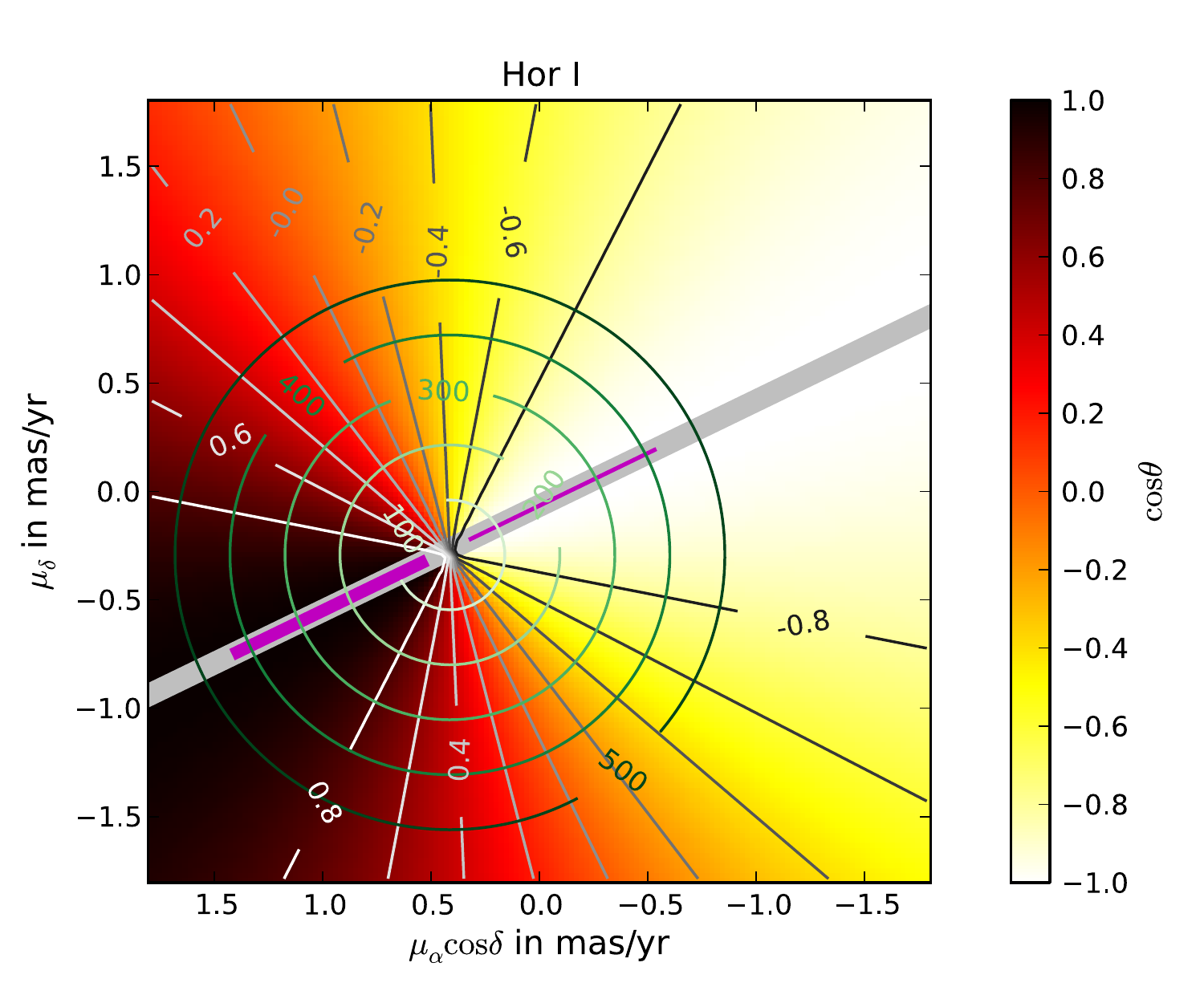}
\includegraphics[width=58mm]{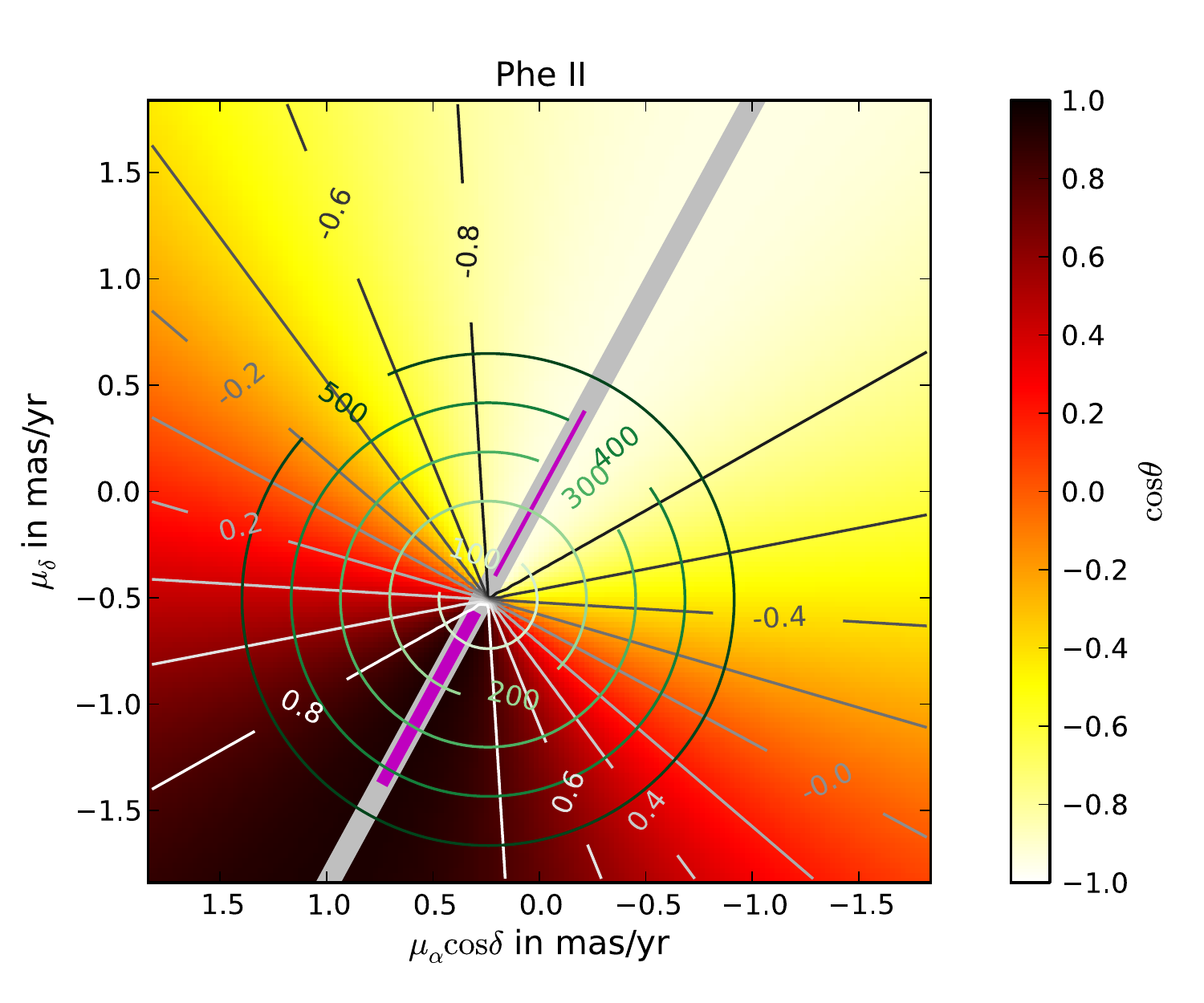}
\includegraphics[width=58mm]{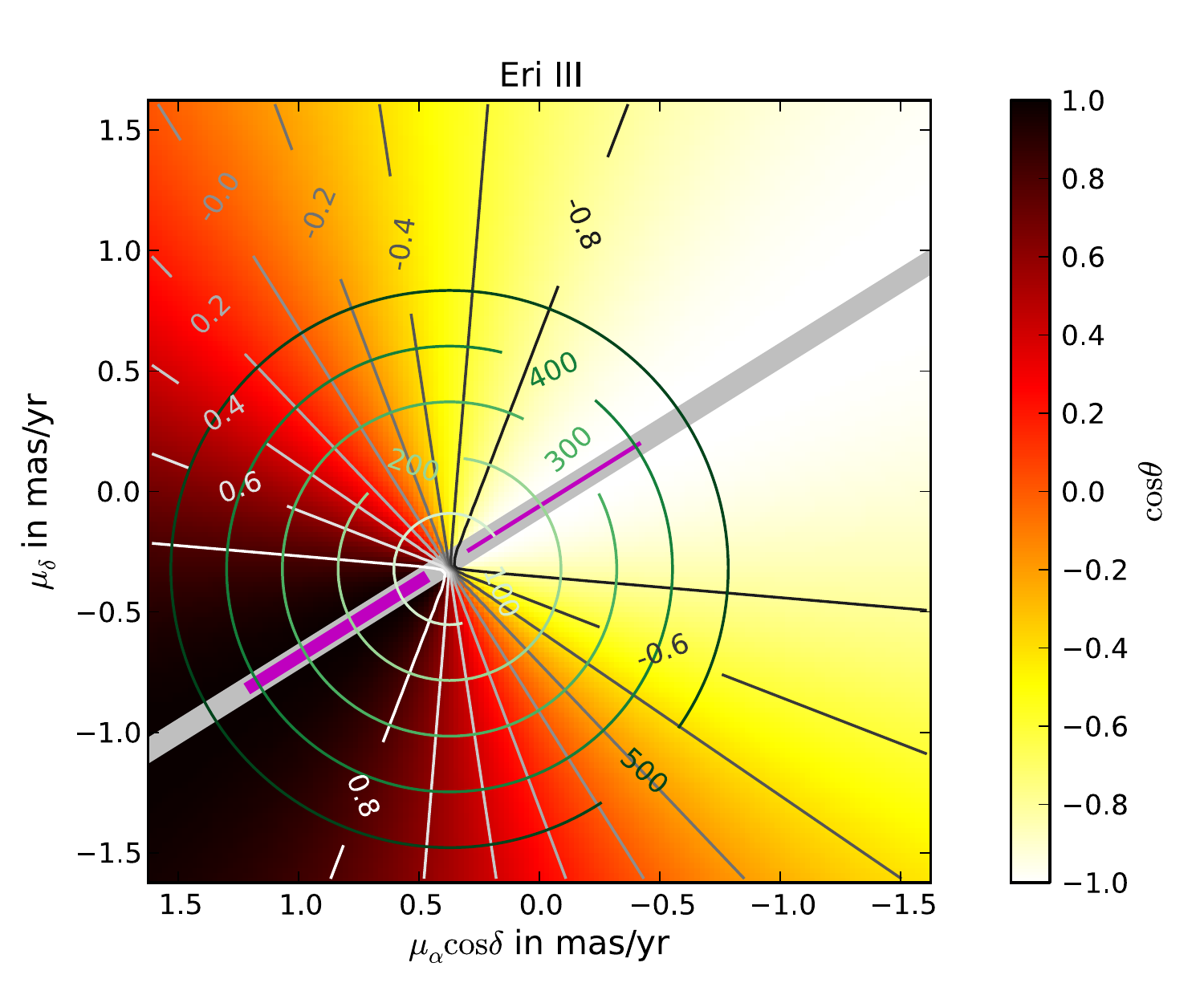}
\includegraphics[width=58mm]{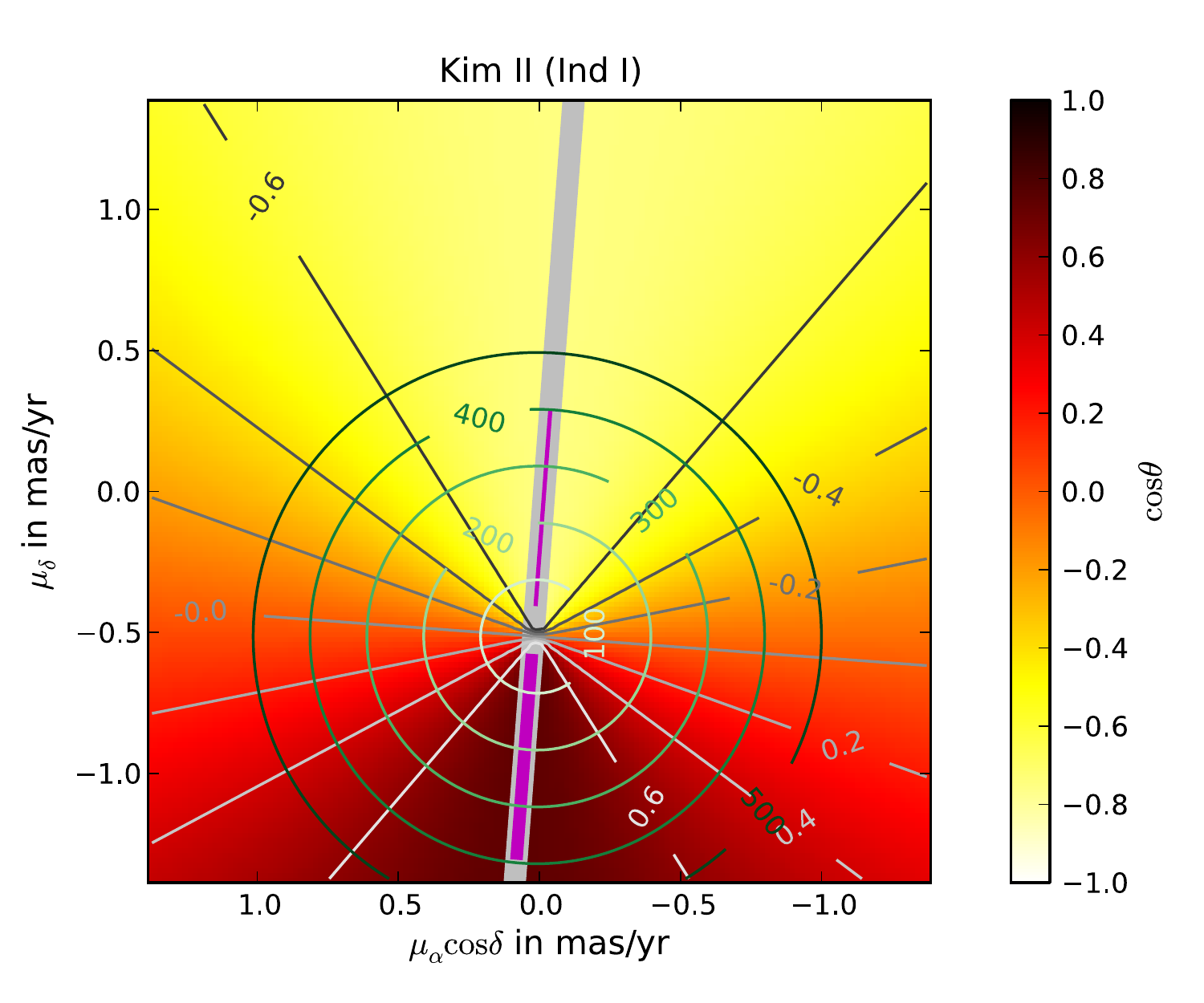}
\includegraphics[width=58mm]{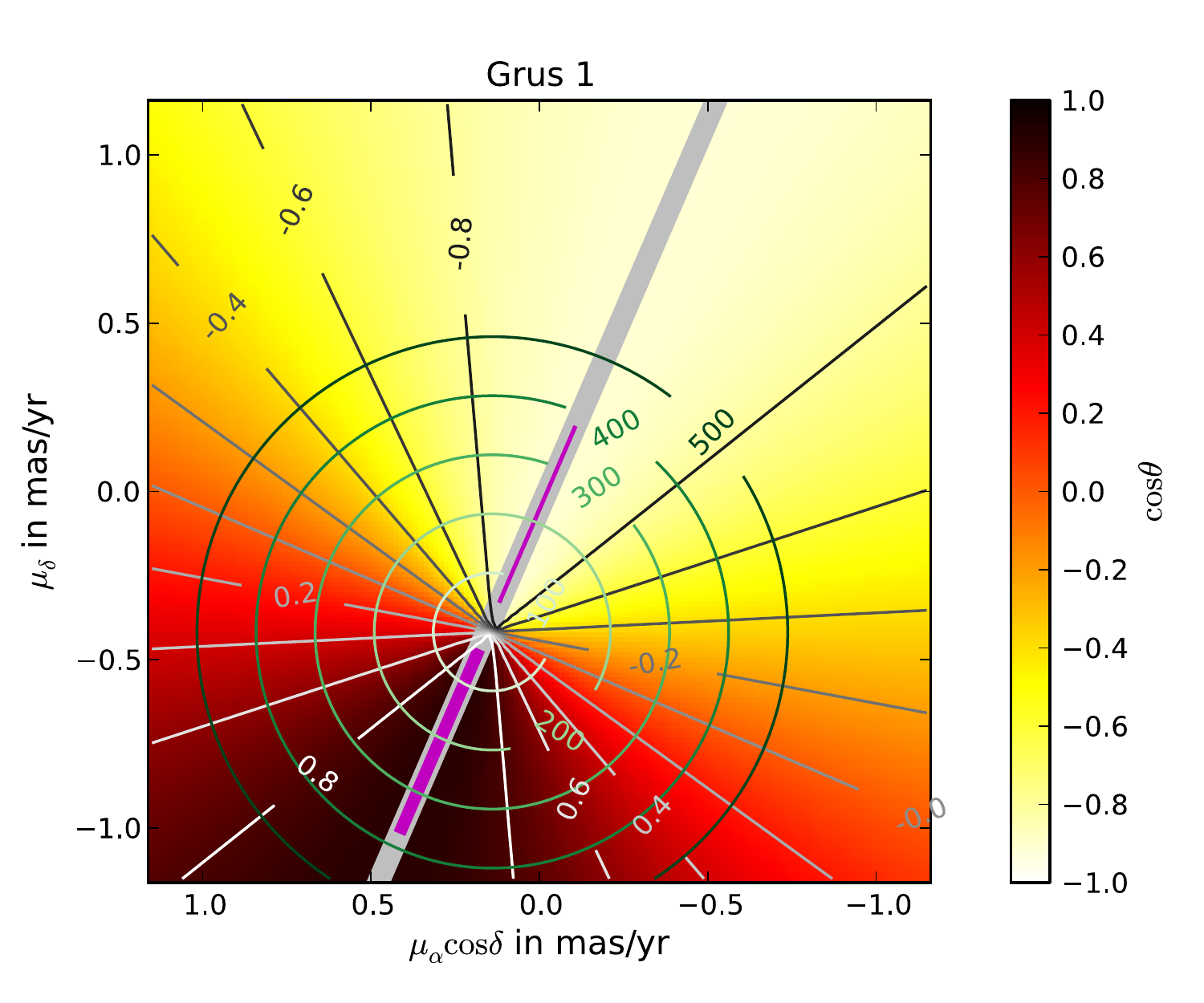}
\includegraphics[width=58mm]{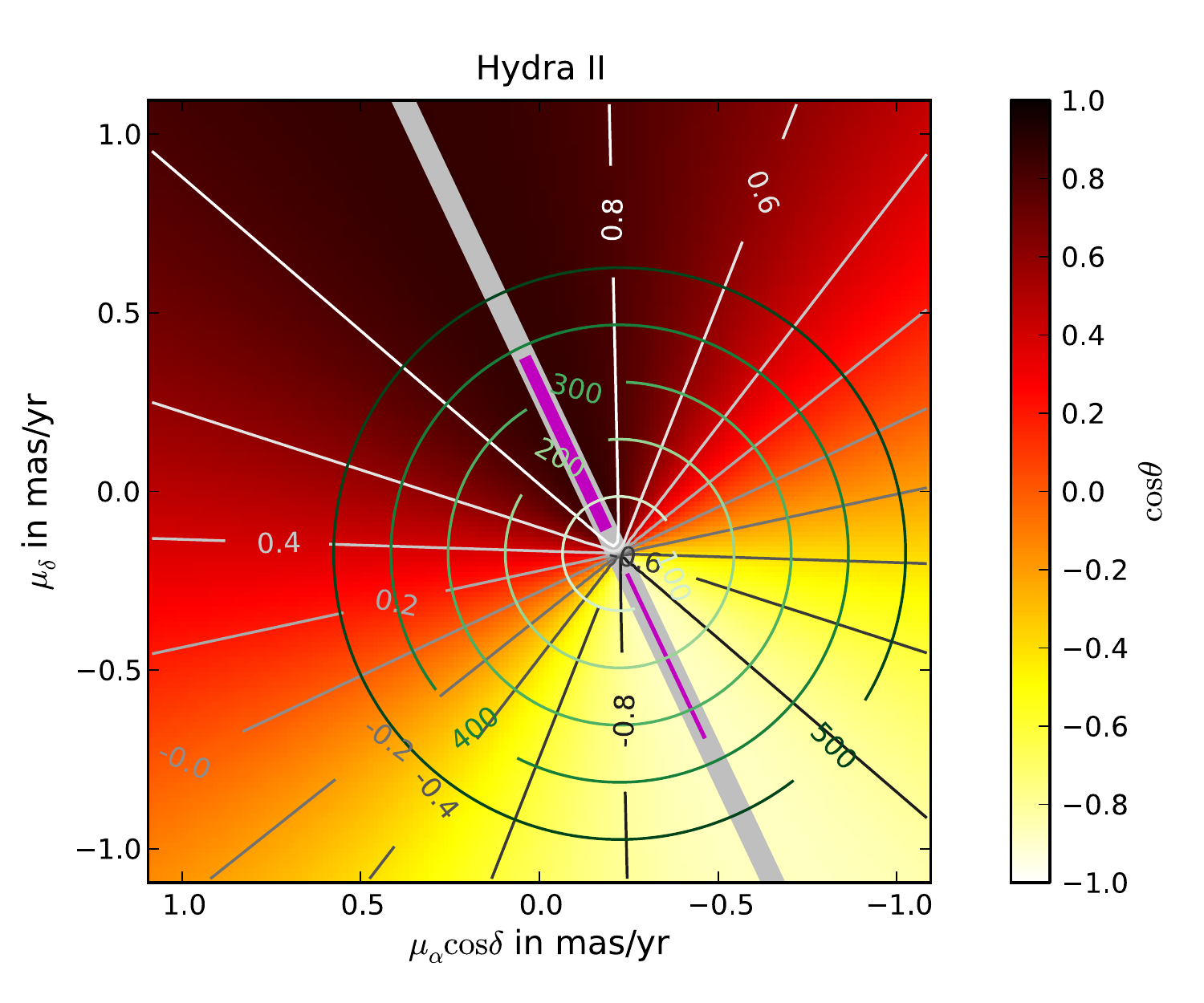}
\includegraphics[width=58mm]{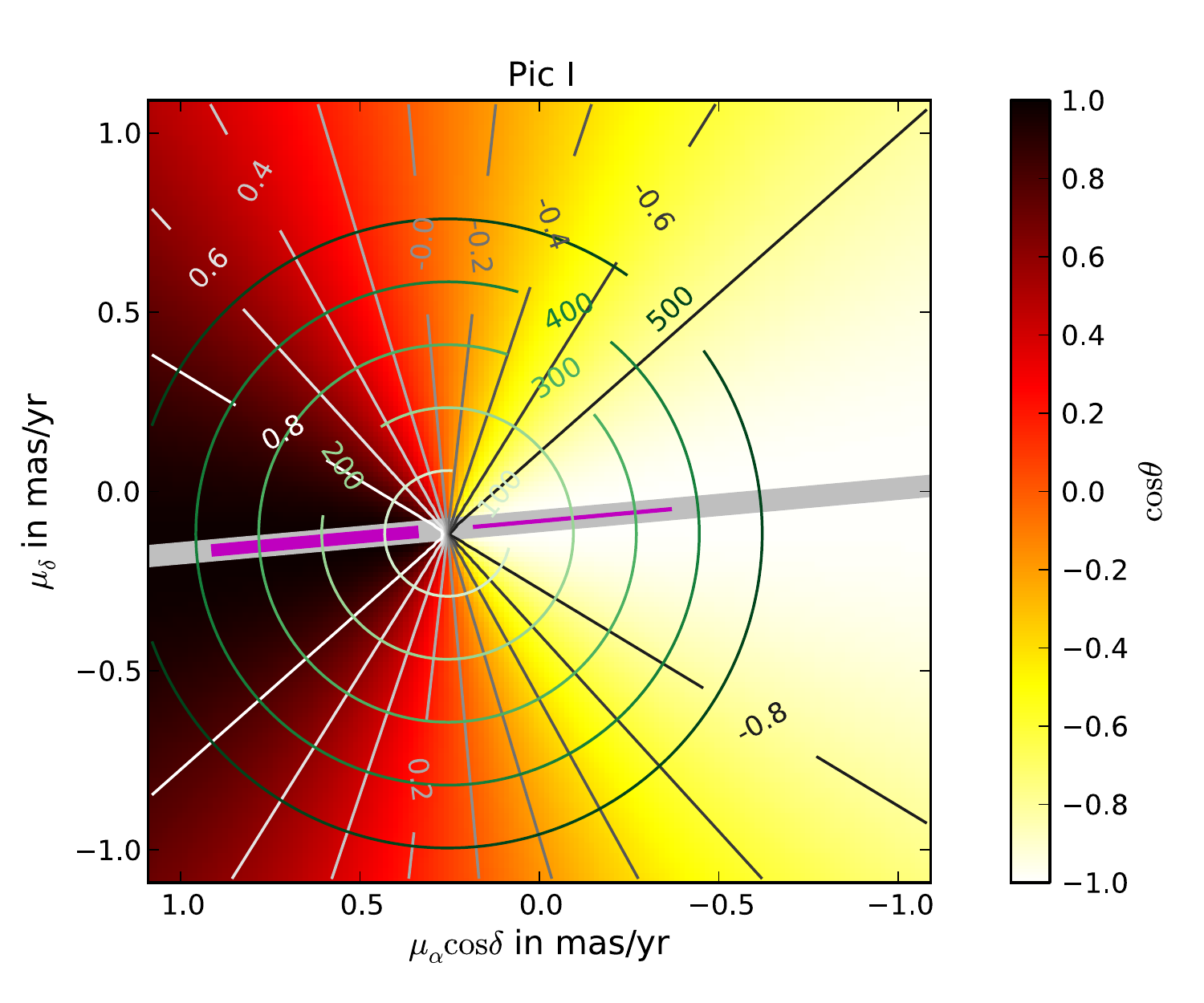}
\includegraphics[width=58mm]{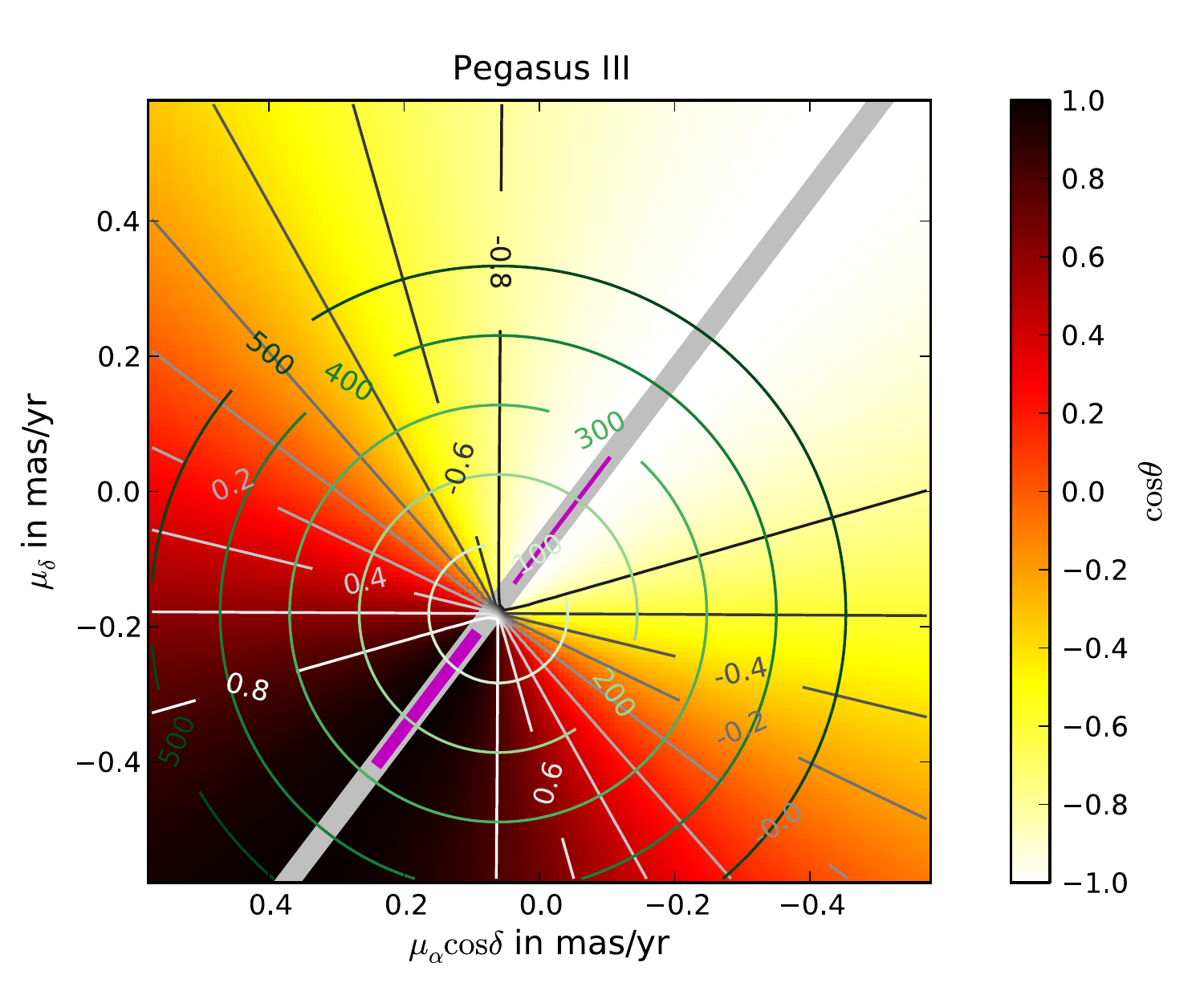}
   \caption{Predicted proper motions of the new MW satellite objects assuming that they move within the VPOS-3 (which is closely aligned with the orbital plane of the LMC). 
The map illustrates the cosine of the angle $\theta$\ between the VPOS-3 plane (the plane fitted to all confirmed satellite galaxies except three outliers, see \citealt{Pawlowski2013} for more details) and the current orbital plane which would result from the PM for each combination of the components $\mu_{\alpha} \cos \delta$\ and $\mu_{\delta}$. The radial grey contour lines illustrate $\cos \theta$\ in steps of 0.2. For $\cos \theta > 0.8$\ an object's orbital plane is inclined by less than $37^{\circ}$\ with the VPOS-3, and can therefore be considered to be co-orbiting (best co-orbiting alignment is marked with a thick magenta line). For $\cos \theta < -0.8$\ it would be counter-orbiting (best alignment marked with a thin magenta line). The green circular contours indicate the lower limits on the absolute speed of the objects relative to the MW in $\mathrm{km\,s}^{-1}$. Since their kinematics have not yet been measured, this assumes that they have zero line-of-sigh velocity relative to the Galactic standard of rest, resulting in the Heliocentric line-of-sight velocities as listed in Table \ref{tab:propmo}. 
 }
 \label{fig:propmo}
\end{figure*}

Out of the 11 MW satellite galaxies for which proper motions (PMs) have been measured, eight are consistent with co-orbiting in the same sense within the VPOS, while Sculptor also orbits within the VPOS but in the opposite direction \citep{Metz2008,PawlowskiKroupa2013}. This orbital alignment finds further support in the alignment of several streams in the MW halo with the VPOS \citep{Pawlowski2012}, most prominently the Magellanic Stream. That satellite galaxy planes appear to be rotating structures is also supported by the M31 satellite plane which shows coherent line-of-sight velocities indicative of rotation \citep{Ibata2013}. In addition, the unexpected anti-correlation of velocities for diametrically opposite satellite galaxy pairs in the SDSS indicates that co-orbiting satellite galaxy planes might be ubiquitous \citep{Ibata2014}.

Since most of the newly discovered objects are closely aligned with the VPOS-3 plane and thus with the average orbital plane of the co-orbiting MW satellite galaxies, we can empirically predict the PM of the new MW satellite objects by using the method presented in \citet{PawlowskiKroupa2013}. This method assumes that the object it either co- or counter-orbiting in the plane defined by the current satellite positions. Because no line-of-sight velocity has yet been measured for most of the new objects, we assume that they have minimum (i.e. zero) Galactocentric line-of-sight velocities $\min(v_{\mathrm{los}})$, calculated as the negative of the component of the Galactocentric Solar motion in the direction of the object. The exact line-of-sight velocity of an object does not affect the orientation of the predicted PM, but only constrains the maximum and minimum predicted PM to values such that it remains bound to the MW. By assuming a minimum Galactocentric line-of-sight velocity the predicted PM range will be maximal. Once line-of-sight velocities have been measured they can be used to put tighter constraints on the expected range in PMs.

To make the prediction compatible with those presented in \citet{PawlowskiKroupa2013} and \citet{PawlowskiKroupa2014}, we adopt the same values for the circular velocity of the MW, the solar motion with respect to the local standard of rest and the Galactocentric distance of the Sun. This is also one reason why the orientation of the VPOS-3 was used in this prediction instead of the new fit. Another reason is that it is not yet known with certainty which of the new objects are satellite galaxies and which are star clusters that should not be part of the VPOS fit. However, since the measured VPOS orientation changes only minimally if the new objects are included in the plane fits, this choice does not strongly affect the predicted PMs anyway. We again use the average of the distances reported by K15 and B15 for those objects discovered by both, except for Kim\,2 for which we use the more reliable measurement from \citet{Kim2015} based on follow-up observations. We also use the method to predict Eri II's PM, but caution that its large distance from the MW makes it unlikely that it is a satellite gravitationally bound to the MW.

The resulting PM predictions are compiled in Table \ref{tab:propmo} and illustrated in Figure \ref{fig:propmo}. The table also provides the angle between the Galactocentric position of the object and the VPOS-3 plane, $\theta_{\mathrm{VPOS-3}}^{\mathrm{predicted}}$, which is equal to the minimum possible inclination between the orbit of the object and the VPOS-3 plane. These angles tend to be small, again illustrating that most of the new objects are closely aligned with the satellite plane.

If future proper motion measurements reveal that more MW satellites follow the strong velocity alignment of the classical MW satellites with the VPOS, this would dramatically increase the significance of their alignment. While the probability of a position drawn from an isotropic distribution to be within an angle $\theta$\ from a given plane is $P_{\mathrm{vector}} = \sin(\theta)$, the probability that a randomly oriented orbital plane is aligned to within $\theta$\ with a given plane is only $P_{\mathrm{plane}} = 1 - \cos(\theta)$. For example, while 71\,per cent of isotropically distributed positions are expected to lie within $45^{\circ}$\ of a randomly oriented plane, only 29\,per cent of isotropically distributed orbital poles are expected to be within $45^{\circ}$\ of the plane's normal vector.

\subsection{Interpreting PM predictions}

We caution against interpreting the PM predictions too tightly. The predicted PM ranges are derived from assuming the best possible alignment of the orbital pole with the normal to a satellite plane. Because the VPOS has a finite rms height, the orbits of most satellites can not be perfectly aligned. The orbital poles of the MW satellites therefore have a minimum intrinsic scatter around their average direction. In addition, all satellites which have orbital planes inclined from the VPOS will at some point during their orbits be perfectly aligned with the VPOS. If observed at that position we would predict the satellites to orbit perfectly within the VPOS, even though they do not.

The width of the VPOS can give us an idea of this intrinsic scatter in orbital pole directions (to which scatter due to PM measurement uncertainties will be added). The VPOS-3 which is used for the PM prediction has a rms axis ratio of $c/a = 0.2$. This gives an approximate opening angle\footnote{Strictly speaking the angle depends on (and decreases with) the radial distance from the MW, because the VPOS is a plane-like structure.} of $2 \times \arcsin\left(0.2\right) \approx 23^{\circ}$. This estimate agrees with the measured spherical standard deviation of the eight best-aligned orbital poles derived from the observed PMs, which is about $27^{\circ}$. To meaningfully compare the predicted and observed PMs therefore requires that we measure the angle between the orbital poles (or the measured orbital pole and the VPOS normal vector assumed for the prediction). If the inclination is less than the expected scatter, the measurement can be said to agree with the prediction.

\section{Predicted velocity dispersions}
\label{sect:veldisp}

\begin{table*}
\begin{minipage}{180mm}
 \caption{Predicted velocity dispersions for each set of photometric parameters compiled in Table \ref{tab:obsprop}}
 \label{tab:veldisp}
 \begin{center}
 \begin{tabular}{@{}lccccccc}
 \hline 
Name & $\sigma_{\mathrm{Newton}}$ & $\sigma_{\mathrm{M07}}$ & $\sigma_{\mathrm{W09}}$ & $\sigma_{\mathrm{MOND}}$ & $\frac{\sigma_{\mathrm{MOND}}}{\sigma_{\mathrm{Newton}}}$ & $a_{\mathrm{in}}$ & $a_{\mathrm{ext}}$ \\
& \multicolumn{4}{c}{[km s$^{-1}$]} &  & \multicolumn{2}{c}{[km$^{2}$\,s$^{-2}$\,kpc$^{-1}$]} \\
 \hline
Kim 1   &  0.14 & 1.4 & 1.2 & 0.18 & 1.3 & 110 & 2280\smallskip\\
Ret II  &  0.38 & 3.2 & 1.6 & 0.68 & 1.8 & 120 & 1160\\
        &  0.26 & 3.0 & 1.6 & 0.46 & 1.8 &  90 & 1210\smallskip\\
Tri II  &  0.17 & 3.1 & 1.6 & 0.32 & 1.9 &  60 & 1060\smallskip\\
Tuc II  &  0.24 & 5.9 & 2.1 & 0.56 & 2.4 &  40 &  660\\
        &  0.23 & 7.6 & 2.3 & 0.61 & 2.6 &  30 &  540\smallskip\\
Hor II  &  0.21 & 3.7 & 1.7 & 0.61 & 2.9 &  60 &  430\\
Hor I   &  0.28 & 4.2 & 1.8 & 0.86 & 3.1 &  70 &  390\\
        &  0.38 & 2.9 & 1.6 & 1.11 & 2.9 & 130 &  430\smallskip\\
Phe II  &  0.41 & 3.1 & 1.6 & 1.30 & 3.2 & 140 &  370\\
        &  0.30 & 2.8 & 1.5 & 0.89 & 3.0 & 110 &  430\smallskip\\
Eri III &  0.39 & 1.8 & 1.3 & 1.27 & 3.2 & 230 &  350\\
        &  0.25 & 2.3 & 1.4 & 0.78 & 3.1 & 120 &  390\smallskip\\
Kim 2  &  0.24 & 1.9 & 1.3 & 0.79 & 3.3 & 130 &  350\\
        &  0.34 & 1.9 & 1.3 & 0.89 & 2.6 & 190 &  550\\
        &  0.34 & 3.4 & 1.7 & 1.10 & 3.2 & 110 &  360\smallskip\\
Gru I   &  0.25 & 4.5 & 1.9 & 0.88 & 3.6 &  60 &  290\smallskip\\
Pic I   &  0.36 & 3.5 & 1.7 & 1.35 & 3.7 & 110 &  270\\
        &  0.32 & 3.0 & 1.6 & 1.15 & 3.6 & 110 &  290\smallskip\\
Hyd II  &  0.47 & 4.4 & 1.9 & 1.78 & 3.8 & 110 &  270\smallskip\\
Peg III &  0.27 & 4.7 & 1.9 & 1.30 & 4.8 &  70 &  160\smallskip\\
Eri II\footnote{Eri II is the only object in the isolated MOND regime, i.e. for which $a_{\mathrm{ext}} < a_{\mathrm{in}}$.} & 1.00 & 6.7 & 2.2 & 3.3 & 3.3 & 160 & 100\\
        &  0.68 & 7.0 & 2.2 & 2.8 & 4.1 & 100 &   90\\
         \hline
 \end{tabular}
 \end{center}
 \small \medskip
$\sigma_{\mathrm{Newton}}$: Predicted velocity dispersion assuming $M/L = 2$\ and that the objects are dark-matter-free.\\
$\sigma_{\mathrm{M07}}$: Predicted velocity dispersion assuming the empirical dark matter halo scaling relation of \citet{McGaugh2007}.\\
$\sigma_{\mathrm{W09}}$: Predicted velocity dispersion assuming the empirical dark matter halo scaling relation of \citet{Walker2009}.\\
$\sigma_{\mathrm{MOND}}$: Predicted MONDian velocity dispersions, assuming stellar mass-to-light ratios of $M/L = 2$\ for all satellite objects.\\
 $\sigma_{\mathrm{MOND}}/\sigma_{\mathrm{Newton}}$: Ratio of predicted MONDian to predicted Newtonian velocity dispersion.\\
 $a_{\mathrm{in}}$: Internal MONDian acceleration of the satellite at its half-light radius.\\
 $a_{\mathrm{ext}}$: External acceleration acting on the satellite due to the potential of the MW.
\end{minipage}
\end{table*}

Here we attempt to predict the internal velocity dispersions of the newly discovered satellites from their reported photometric properties. Since it is unclear whether some of these objects are dwarf satellite galaxies or star clusters, we consider both possibilities.  We consider both conventional gravity and MOND \citep{Milgrom1983}. Since differences in the photometric properties reported for the objects translate into differences in the predicted velocity dispersions, we make predictions using each of the reported sets of properties.  Our results are compiled in Table \ref{tab:veldisp} in the order in which they are listed in Table \ref{tab:obsprop}.

\subsection{Star Clusters}

If the newly discovered objects are star clusters, then they may be devoid of dark matter. In this case, their velocity dispersions follow directly from the virial relation and the observed stellar mass and half-light radii:
\begin{equation}
\sigma \approx \left(\frac{GM_*}{3r_{1/2}}\right)^{1/2}
\end{equation}
where $r_{1/2}$ is the 3D half light radius (estimated as 4/3 the effective radius) and $M_* = \Upsilon_* L_V$. To estimate the stellar mass, we assume a mass-to-light ratio of $\Upsilon_* = 2\;M_{\sun}/L_{\sun}$. This is extremely uncertain (see \S \ref{sec:SM}). The velocity dispersions predicted in this way are given in Table~\ref{tab:veldisp}. These are low luminosity systems, so the predicted velocity dispersions are small: $< 1\;\kms$ in all cases, and $< 0.5\;\kms$ in all but one case.
However, the disruption of such dark matter free systems on their orbits around the Milky Way can increase their \textit{apparent} mass-to-light ratios substantially \citep{Kroupa1997}. This successfully predicted an object like the later discovered MW satellite Hercules \citep[see the discussion in][]{Kroupa2010}.

\subsection{Scaling Relations}

If the newly discovered objects are dwarf satellite galaxies, then we expect them to reside within dark matter sub-halos.  If this is the case, the kinematics are presumably dominated by dark matter as with the other known dwarfs.  We then anticipate higher velocity dispersions.

There is no universally agreed method to predict the velocity dispersions of individual dwarf satellite galaxies in $\Lambda$CDM. The correlation between luminosity and halo mass is exceedingly weak on these scales \citep{Wolf2010}, while the expected scatter in  $\Lambda$CDM sub-halo properties is large \citep{Tollerud2011}. Consequently, the observed luminosity should have little predictive power, with the velocity dispersions being essentially stochastic.  

There are empirical scaling relations that we can use to anticipate the velocity dispersion of the newly discovered objects. This is done with the scaling relations of \citet{McGaugh2007} and \cite{Walker2009}, the results of which are tabulated in Table~\ref{tab:veldisp}. \cite{Walker2010} showed that the distinct relations of \citet{McGaugh2007} and \cite{Walker2009} were consistent with each other and the data available at the time. However, they are not identical, and give different results when extrapolated into the regime represented by the new objects. We therefore tabulate distinct predictions for each, bearing in mind that these are the extrapolations of empirical scaling relations and are not predictions derived from a specific theory like $\Lambda$CDM.

\citet{McGaugh2007} fit the baryon subtracted rotation curves of spiral galaxies to obtain an  expression for the rotation velocity due to the dark halo component. Though fit to spirals at larger radii, this relation did a good job\footnote{Evaluation of the scaling relation of \citet{McGaugh2007} at 300 pc anticipates $M(< 300\;\mathrm{pc}) = 1.8 \times 10^7\;\mathrm{M}_{\sun}$.} of anticipating the  common mass scale found for satellite galaxies by \citet{Strigari2008}. Assuming $\sigma = V_h/\sqrt{3}$ and evaluating at the observed half light radius (in kpc), the relation of \citet{McGaugh2007} becomes
\begin{equation}
\log \sigma = 1.23+0.5 \log r_e.
\label{eq:M07scaling}
\end{equation}
This anticipates velocity dispersions in the range of 1 -- $7\;\kms$ (Table~\ref{tab:veldisp}).

\citet{Walker2009} fit the data for dwarf satellite galaxies to obtain
\begin{equation}
\log \sigma \approx 0.5 + 0.2 \log r_e.
\end{equation}
This should be more applicable to the newly discovered objects, if they are indeed dwarf satellites, albeit very small ones. The dependence on size in this regime is rather weaker, anticipating velocity dispersions in the range 1 -- $2.3\;\kms$ (Table~\ref{tab:veldisp}). Note that both scaling relations are known to be violated in some cases \citep{Collins2014}.

We may of course have a heterogeneous mix of objects: some might be dwarf satellite galaxies in sub-halos, while others might simply be star clusters.  If so, the different anticipated velocity dispersions should help distinguish these two cases.  However, considerable observational care will be required to do so given the small anticipated dispersions.

\subsection{MOND}

We also tabulate the velocity dispersions predicted by MOND \citep{Milgrom1983}. In this theory, the velocity dispersion should follow from the observed properties of each object, irrespective of whether it is a star cluster or dwarf satellite galaxy.  Either way, the physics is the same.

To predict velocity dispersions with MOND, we follow the procedure outlined by \citet{McGaughMilgrom2013a}. We assume (as done there) that $\Upsilon_* = 2\;M_{\sun}/L_{\sun}$. This approach has had considerable success in predicting, often \textit{a priori}, the velocity dispersions of the dwarf satellites of Andromeda and of isolated dwarfs in the LG \citep{McGaughMilgrom2013b,PawlowskiMcGaugh2014a}.
A different method employed by \citet{Lueghausen2014} has also been successful in predicting the velocity dispersions of the most luminous MW dSph satellites Fornax and Sculptor, but found the measured velocity dispersions of Sextans, Carina and Draco to be higher than predicted.

The velocity dispersion predicted by MOND depends on whether the internal gravitational field $a_{in}$ of an object dominates (the isolated case), or if it is dominated by the external field effect (EFE) of the host galaxy $a_{ex}$. The case that applies depends on the relative strength of the internal and external fields: a system is considered to be in the MOND regime and isolated if $a_{ex} < a_{in} < a_0$, and is in the EFE regime if $a_{in} < a_{ex}$. There are thus two velocity dispersion estimators: that for the isolated case \citep[eq.~2 of][]{McGaughMilgrom2013a}, and that for the EFE case \citep[eq.~3 of][]{McGaughMilgrom2013a}. 

The internal field depends only on the properties of each object, and is estimated at the half light radius as in \citet{McGaughMilgrom2013a}. The external field depends on the total baryonic mass of the Milky Way.  For specificity, we adopt the empirical (``bumps \& wiggles'') Milky Way model of \citet{McGaugh2008}, which provides an estimate of $a_{ex} = V_{MW}^2/R$ at the galactocentric distance $R = d_{MW}$ of each object.  The uncertainty in the circular velocity of the Milky Way at these distances affects the predicted velocity dispersions at the $\pm 0.3\;\kms$ level.

The EFE dominates in all but the most distant case: Eri II is the only object in Table~\ref{tab:veldisp} in the isolated regime.   In some cases, the external field is only marginally dominant.  In these cases, neither mass estimator is really adequate, and the velocity dispersion may be slightly under-predicted \citep{Milgrom1995}. In all cases, the enhancement in the velocity dispersion predicted by MOND is rather modest. Typically it is only a factor of a few above the purely Newtonian (star cluster) case, and is often less than anticipated by the dark halo scaling relations. The many uncertainties of the quantities that go into the prediction are compounded by the potential for systematic errors.

\subsection{Challenges to Interpretation}

Can we hope to observationally distinguish between the various predictions?
An accuracy of $\sim 0.1\;\kms$ is required to resolve the velocity dispersions anticipated for star clusters.
This is certainly possible, if challenging.  

The larger concern is systematic uncertainties. For example, are these tiny systems in dynamical equilibrium?
If not, we might misinterpret a high velocity dispersion of a dissolving star cluster as the equilibrium dispersion
of a dwarf satellite residing in a dark matter sub-halo.

Many of the uncertainties involved in measuring and interpreting velocity dispersions have been discussed 
by \citet{McGaughWolf2010}.  
We emphasize here just two possible systematics that we fear will make it extremely difficult to distinguish between
the various possibilities.  One, binary stars may inflate the observed velocity dispersions.  Two, the conversion from
light to stellar mass is rather fraught for systems containing so few stars.

\subsubsection{Binary Stars}

Some of the individual stars for which velocities are obtained are presumably members of unresolved binaries.
Binary stars are themselves in orbit around one another, possibly at speeds comparable to the velocity dispersion expected
for the system as a whole.  This can inflate the velocity dispersion measured for the system.  
Binaries can easily contribute enough to the measured velocity dispersion to change the interpretation from one extreme to the other.
One can correct for this effect, but it requires the patience of many repeat observations \citep{Simon2011}.
 
\subsubsection{Stellar Mass}
\label{sec:SM}

Another problem when considering ultrafaint dwarfs is the uncertainty in the stellar mass-to-light ratio.
Ultrafaint dwarfs are so small that their entire luminosity can be less than that of a single high mass star.
This violates an essential assumption in the estimation of stellar mass-to-light ratios with stellar population models:
that there are enough stars to statistically sample all phases of stellar evolution.  Indeed, in systems composed
of only a few hundred stars, the evolution of a single star up the giant branch will substantially change the luminosity 
of the entire system without changing its mass.  For this reason, all predictions made with an assumed 
mass-to-light ratio are subject to large uncertainty.  This strongly affects the predictions for both purely Newtonian
star clusters and MOND. 

\subsubsection{Ret II, Hor I and Hyd II}

\begin{figure}
 \centering
 \includegraphics[width=80mm]{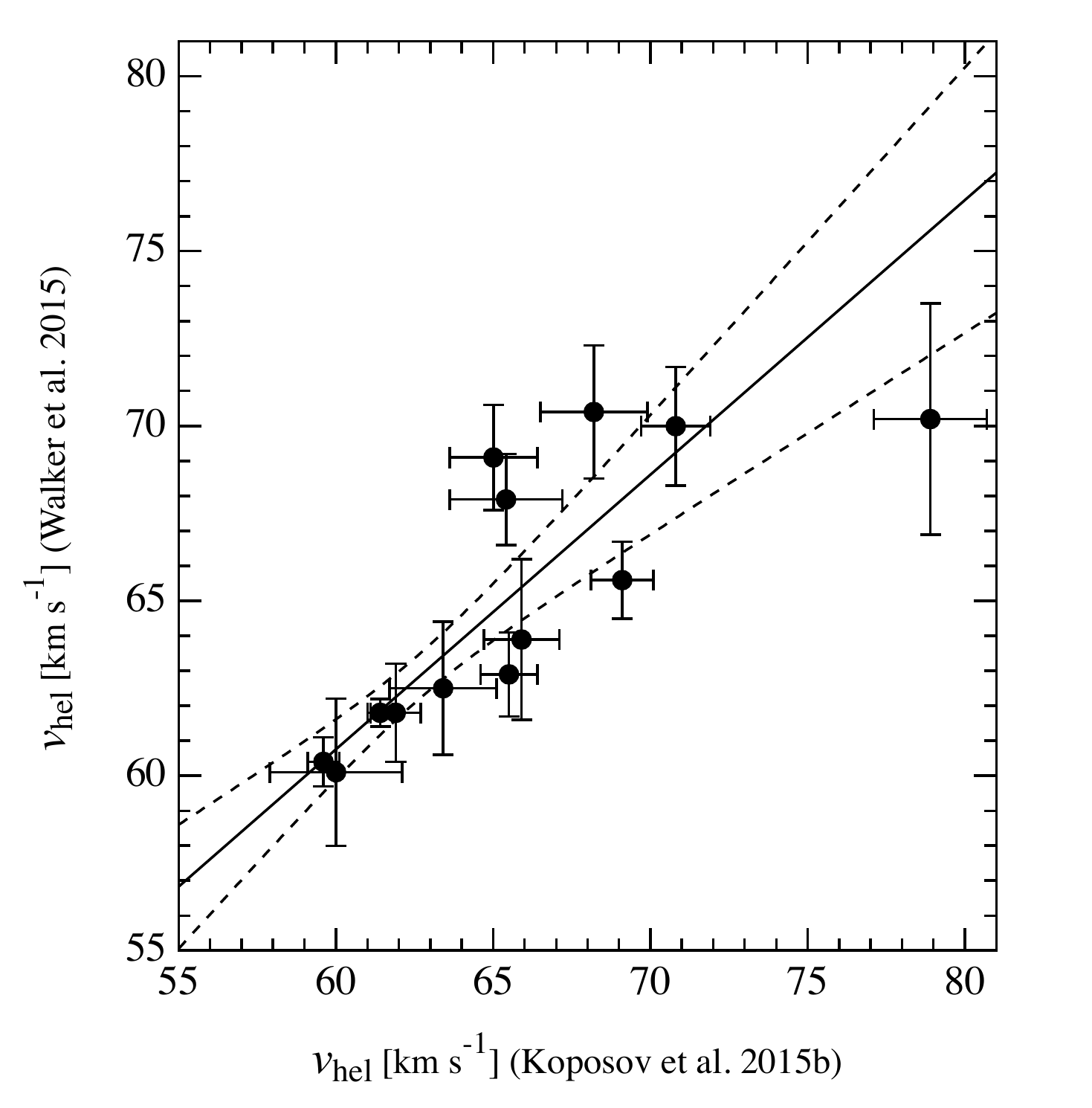}
 \caption{
Velocity measurements for those Ret\,II stars in common between the sets \citet{Walker2015}, plotted on the vertical axis, and \citet{Koposov2015b}, plotted on the horizontal axis. The data points are plotted with their respective error bars, also shown is the best fitting line and the upper and lower 2-sigma confidence bands. We find that the measurements agree well with each other, giving confidence in the reported heliocentric velocity and velocity dispersion for this object.
 }
 \label{fig:velcompare}
\end{figure}

Ret\,II is the first of the objects in Table~\ref{tab:veldisp} to have a measured velocity dispersion $\sigma$\ and heliocentric velocity $v_{\mathrm{hel}}$. \citet{Walker2015} measure $\sigma = 3.6^{+0.9}_{-0.6}\;\kms$\ ($v_{\mathrm{hel}} = 64.8^{+1.1}_{-1.0}\,\mathrm{km\,s}^{-1}$), \citet{Simon2015} measure $\sigma = 3.3\pm0.7\;\kms$\ ($v_{\mathrm{hel}} = 62.8\pm0.5\,\mathrm{km\,s}^{-1}$) and \citet{Koposov2015b} measure $\sigma = 3.2^{+1.6}_{-0.5}\;\kms$\ ($v_{\mathrm{hel}} = 64.7^{+1.3}_{-0.8}\,\mathrm{km\,s}^{-1}$). These measurements are nicely consistent, and are clearly too large for a star cluster devoid of dark matter ($0.3\;\kms$) or for MOND ($0.5\;\kms$). This of course presumes that the system is in dynamical equilibrium, that our guess for the stellar mass-to-light ratio is not far off, and that binary stars contribute $\ll 3\;\kms$ in quadrature to the observed velocity dispersion.

We have used the opportunity provided by the independent measurements of Ret\,II's velocity dispersion to check whether the studies are in mutual agreement or whether systematic errors might be present. This is motivated by the comparison for the Carina dSph by \citet{Godwin1987}.
We can make a similar comparison for the studies of \citet{Walker2015} and \citet{Koposov2015b}, who have 13 Ret\,II stars in common. The respective velocities measured for these stars are plotted against each other in Fig. \ref{fig:velcompare}. We find that the best-fitting line has a slope of $0.79\pm0.21$\ and a y-axis intercept of $13.6\pm13.3$\ (not counting the outlier with the largest error bars which also disagrees with the systematic velocity of Ret\,II by almost $15\;\kms$\ in one of the two studies). Hence, there is indeed a positive correlation between the two velocity sets, which furthermore is consistent with a slope of one. This supports the interpretation that not random errors but indeed the internal velocity dispersion of the object is a major contributor to the spread in velocities.

Taken at face value, the observed velocity dispersion is consistent with the empirical scaling relation of \citet{McGaugh2007}, which anticipates $\sigma = 3$ -- $4\;\kms$, depending on whose photometry  is employed. This implies a dark matter halo consistent with the near-universal halo found by \citet{McGaugh2007} and \citet{Walker2010}. Unfortunately, this empirical dark matter halo is not consistent with $\Lambda$CDM \citep{McGaugh2007}, though presumably it can be accommodated by invoking feedback or some other mechanism.

To make matters worse, the universal halo that is successful in the case of Ret\,II does not work in the cases of the satellites of M31 And XIX, XXI, and XXV \citep{Collins2014}.  These objects are faint, but have much larger effective radii than the objects under consideration here.  Application of equation \ref{eq:M07scaling} of \citet{McGaugh2007} anticipates $\sigma > 13\;\kms$ for these dwarfs of M31, while they are observed to have $\sigma < 5\;\kms$ \citep{Collins2014}.  These objects should be strongly affected by the EFE in MOND, which was unique in accurately predicting their velocity dispersions in advance \citep{McGaughMilgrom2013a,McGaughMilgrom2013b}. We therefore urge caution in interpreting the velocity dispersions of these objects, especially in light of the systematic uncertainties discussed above.

\citet{Koposov2015b} also infer a velocity dispersion for Hor\,I from five stars of $\sigma = 4.9^{+2.8}_{-0.9}\;\kms$\ ($v_{\mathrm{hel}} = 112.8^{+2.5}_{-2.6}\,\mathrm{km\,s}^{-1}$). This dispersion value exceeds all predictions, but it again comes closest to the prediction using the \citet{McGaugh2007} scaling relation. However, for the reasons discussed above one should be extremely cautious in interpreting these velocity dispersion measurements.

After this manuscript was submitted, \citet{Kirby2015} announced the first spectroscopic measurement of stars in Hyd\,II. They did not resolve its velocity dispersion, but report an upper limit of $\sigma < 4.5\,\mathrm{km\,s}^{-1}$\ (95\,per cent confidence), which  is consistent with all predictions. \citet{Kirby2015} found a heliocentric velocity for Hyd\,II of $v_{\mathrm{hel}} = 303.1\pm1.4\,\mathrm{km\,s}^{-1}$, which they report to be similar to the Leading arm of the Magellanic Stream. Such a similarity to the velocity of the Magellanic Stream is a general trend for the objects found to lie within the dwarf galaxy planes in the LG \citep{Pawlowski2013}, which is also followed by Ret\,II and Hor\,I (see Fig. \ref{fig:MagStream}). Unfortunately, the line-of-sight velocity of satellite objects does not provide decisive information on whether it orbits within the VPOS, because this velocity is mostly oriented along the radial component of its position vector from the Galactic center.

\section{Discussion and Conclusions}
\label{sect:conclusions}

\begin{figure*}
 \centering
 \includegraphics[width=180mm]{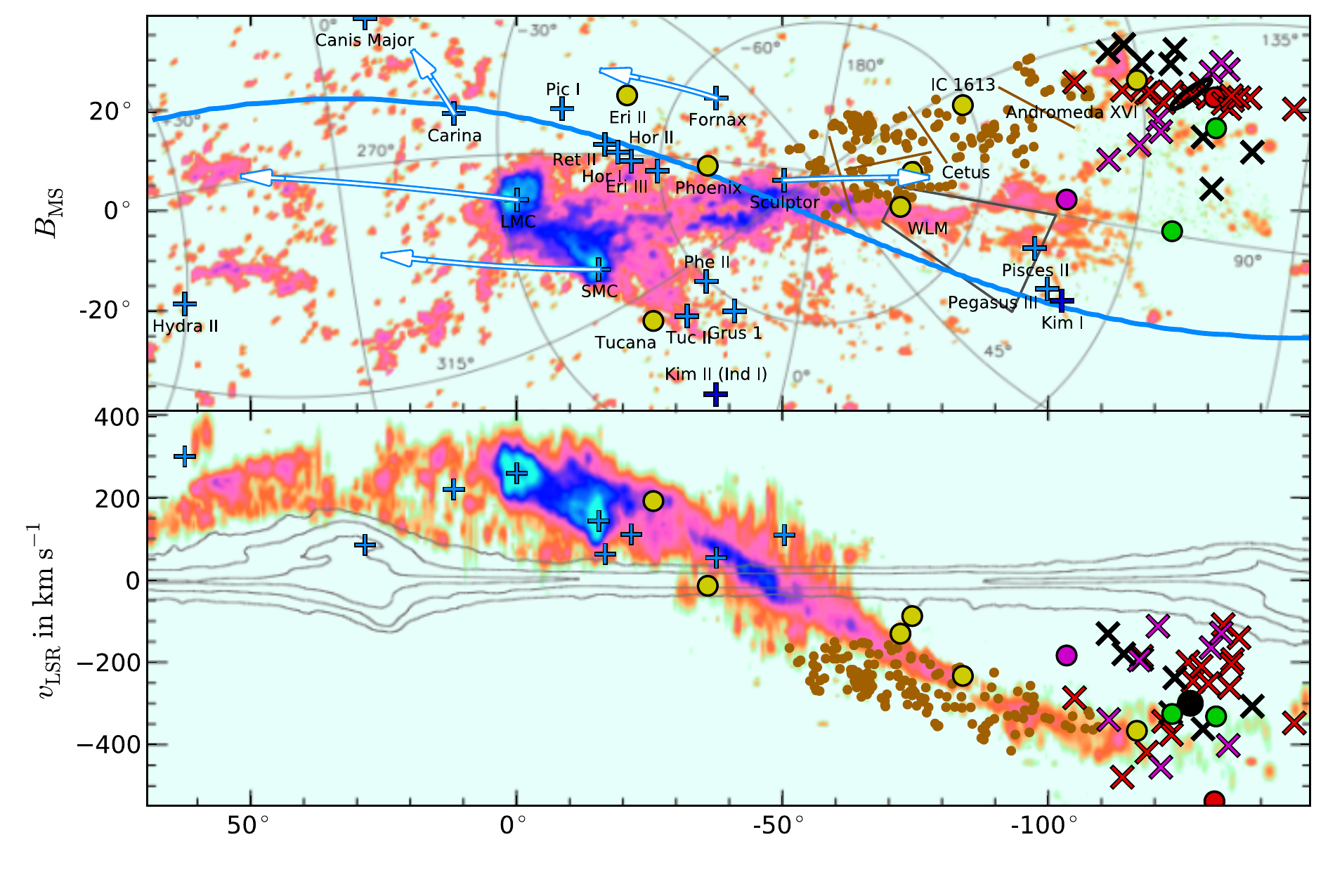}
 \caption{
 Comparison of the Magellanic Stream \citep[from][]{Nidever2010} and the LG dwarf galaxies as in figure 17 of \citet{Pawlowski2013}, but updated to include the newly discovered objects and additional information. 
The \textit{upper panel} plots positions on the sky in the Magellanic Stream Coordinate \citep{Nidever2008}.
 The solid blue line indicates the intersection of the best-fit VPOS plane with a sphere of 250\,kpc radius. The arrows indicate the current most-likely velocity vectors of the MW satellites LMC, SMC, Fornax, Carina and Sculptor, as compiled in table 2 of \citet{PawlowskiKroupa2013}. The velocity vectors are represented in position space by converting $5\,\mathrm{km\,s}^{-1}$\ to length of 1\,kpc and then projected into Magellanic Stream coordinates. As in Fig. \ref{fig:ASP} the velocity vectors reveal that the satellites move mostly along the plane. 
 The blue plus signs indicate the positions of the MW satellites (dark blue for the two star clusters Kim\,1 and 2), yellow dots the positions of the non-satellite dwarf galaxies which are part of a highly flattened plane approximately connecting the MW and M31, and crosses are satellite galaxies of M31 (red for those in the M31 satellite galaxy plane). The black ellipse indicates the position and orientation of M31.  
 The \textit{lower panel} gives the line-of-sight velocities of the Magellanic Stream and those galaxies for which kinematics are available. See \citet{Pawlowski2013} for further details.
 }
 \label{fig:MagStream}
\end{figure*}

We have compiled a list of 14 recently discovered stellar systems in the vicinity of the MW, many of which are probably MW satellite galaxies. We find that most of these objects align well with the VPOS, which consists of both satellite galaxies and star clusters \citep{Pawlowski2012}. The updated VPOS fit parameters, compiled in Table \ref{tab:planeparams}, do not deviate substantially from the previous ones: the rms height is almost unchanged, the orientation preserved to within $9^{\circ}$\ and the offset from the MW center is reduced. Assuming that this alignment indicates the objects to be part of a common dynamical structure, as is indicated by the aligned orbital poles of the 11 classical MW satellites \citep{PawlowskiKroupa2013}, we predict the proper motions of the new satellite objects (see Sect. \ref{sect:propmo}). 

We apply Newtonian and MONDian dynamics and different dark matter halo scaling relations to predict the velocity dispersions of the objects from their photometric properties (see Sect. \ref{sect:veldisp}). These three distinct assumptions result in predictions with only modest differences. For most objects Newtonian dynamics predict velocity dispersion between 0.2 and $0.4\,\mathrm{km\,s^{-1}}$, the dark matter scaling relations predict velocity dispersions between 1 and $4\,\mathrm{km\,s^{-1}}$, and the MOND predictions lie in between these. This small range of very low velocity dispersions makes it extremely difficult to discriminate between the three cases observationally, which will require very precise measurement and good control of systematic effects such as unresolved binary stars.

Most of the objects are in the southern hemisphere of the MW, and the majority of them have been discovered in the DES. The survey footprint lies close to the Magellanic clouds which orbit within the VPOS, such that an alignment with the VPOS might not be unexpected. However, even though the area covered by the DES so far falls close to the VPOS, one would expect the ten objects discovered in the data to have about 50\,per cent larger mean and median offsets from the VPOS if they were drawn from an isotropic distribution confined to the survey footprint. Both mean and median offsets at least as small as observed are rare among such randomized realisations (9 and 4 per cent, respectively). Furthermore, several other objects were discovered elsewhere, away from known concentrations of satellite galaxies, but nevertheless aligned with the VPOS. The fact that the PanSTARRs survey, despite its $3\pi$\ sky coverage, has so far not resulted in the discovery of a large number satellite galaxies outside of the VPOS provides further hints that even the fainter MW satellites align with the satellite structure which was first discussed almost 40 years ago by \citet{KunkelDemers1976} and \citet{LyndenBell1976}. 

Among the M31 satellite galaxies a similar, and apparently also co-rotating, plane consisting of about half of the satellite population was found by \citet{Ibata2013}. It is aligned with the M31's prominent stellar streams \citep{Hammer2013}, which is reminiscent of the preferential alignment of streams in the MW halo with the VPOS, most prominently the Magellanic Stream. Co-orbiting planes and similar satellite alignments might even be common throughout the Universe \citep{Ibata2014,PawlowskiKroupa2014,Tully2015}.

The new satellites could prove to be important for the wider picture of dwarf galaxies and the overall dynamics in the Local Group (LG). The non-satellite LG dwarf galaxies are confined to two highly symmetric and extremely narrow planes \citep{Pawlowski2013}. The dominant of these two planes appears to connect M31 and its satellite galaxy plane with the VPOS around the MW, and agrees in projected position and line-of-sight velocity with the Magellanic Stream (see section 7.4 in \citealt{Pawlowski2013} and Fig. \ref{fig:MagStream}). The new discoveries are particularly interesting because many lie in the vicinity of the Magellanic Clouds but into the direction of M31, close to the Magellanic Stream. This region was identified as the ``direction of decision'' by \citet{Pawlowski2013}, because it is where the MW and M31 satellite planes intersect with the dominant plane of non-satellite dwarf galaxies in the LG. Knowledge of the phase-space distribution of objects in this region should help to determine if and how these structures are connected.

\section*{Acknowledgements}

We thank Benoit Famaey, Pavel Kroupa, Federico Lelli, Mario Mateo and Matthew Walker for useful discussions and comments. HJ acknowledges the support of the Australian Research Council through Discovery Project DP150100862.The contributions of MSP and SSM to this publication were made possible through the support of a grant from the John Templeton Foundation. The opinions expressed in this publication are those of the author and do not necessarily reflect the views of the John Templeton Foundation.

\bibliographystyle{mn2eB}
\bibliography{./VPOSnewref}

\label{lastpage}

\end{document}